\documentstyle[multicol,pre,aps,epsf]{revtex}

\newcommand{\be}{\begin{eqnarray}}
\newcommand{\ee}{\end{eqnarray}}
\newcommand{\vonK}{von K\'arm\'an }

\begin{document}
\date{\today}
\draft

\title{Properties of Ridges in Elastic Membranes}

\author{Alexander E.~Lobkovsky\footnote{Electronic mail address:
{\tt a-lobkovsky@uchicago.edu}, World Wide Web:
{\tt http://rainbow.uchicago.edu/{\char'176}lobkovsk}.}, T.~A.~Witten}
\address
{The James Franck Institute\\
The University of Chicago\\
5640 South Ellis Avenue
Chicago, Illinois 60637}
\maketitle

\begin{abstract}

When a thin elastic sheet is confined
to a region much smaller than its size
the morphology of the resulting crumpled
membrane is a network of straight ridges or folds
that meet at sharp vertices.
A virial theorem predicts the ratio
of the total bending and stretching energies of a ridge.
Small strains and curvatures persist far away from the
ridge.
We discuss several kinds of perturbations that distinguish
a ridge in a crumpled sheet from an isolated ridge studied
earlier (A.~E. Lobkovsky, Phys. Rev. E. {\bf 53} 3750 (1996)).
Linear response as well as buckling properties are investigated.
We find that quite generally, the energy of a ridge can change
by no more than a finite fraction before it buckles.

\end{abstract}


\begin{multicols}{2}
\section{Introduction}

There is an abundance of phenomena involving strong deformation
of thin elastic membranes that span a wide range of scales.
On the microscopic scale, phospholipid membranes
behave like a solid below a two-dimensional (2D) freezing point
\cite{vesicle}.  Some inorganic compounds such as graphite oxide 
\cite{graphite} and molybdenum disulphite \cite{MoS2} also behave like
elastic membranes on scales large compared to the interatomic
spacing.  The graphite oxide sheets can be collapsed
in solution by inducing an effective attractive interaction between
distant parts of the sheet.  Molybdenum disuphite has also been observed
in a ``rag'' phase that looks similar to a crumpled piece of paper.
Mechanical properties of macroscopic 
thin elastic plates and shells undergoing large deformations are important
in engineering of safety structures \cite{crash} and packaging material
development \cite{cushion}.

Stability and post-bucking properties of thin shells and plates have
been a subject of intense investigation \cite{math-conference}.
Few general results
have been derived however.  Due to the complexity of the equations which
describe large deflections of thin plates (two quadratic fourth order partial
differential equations), rigorous proofs are difficult to achieve.
Difficulties in treating thin shells and plates 
arise due to the fact that a small
parameter related to the thickness of the shell multiplies the highest
derivative term in the equations \cite{shell-theory}.
It is well-known that this fact gives rise to a variety
of boundary layer phenomena in the bending of thin shells \cite{fung}.
Many different types of boundary conditions that lead to a boundary
layer have been analyzed.  They include bending moments \cite{reissner},
shear forces \cite{kelvin} and free boundary conditions with distributed
bending moments \cite{ashwell}.  A common feature that emerges from these
studies is that membrane stresses become confined to the boundary layer
region whose size vanishes as some power of the shell thickness.

It has been suggested recently by the author and others 
\cite{science,vonK} that membrane stresses in a strongly
crumpled elastic sheet are also confined to a set of straight ridge
singularities or folds.
These ridge singularities, which were shown to
arise under quite general conditions, constitute another
case of the boundary layer phenomena in thin plates.
A scaling law for the ridge width as a function of the
plate thickness had been established with the use of an energy
scaling argument \cite{science} and a boundary layer analysis
of \vonK thin plate equations \cite{vonK}.
Elastic energy was found to be confined in the ridges
and to scale as $1/3$ power of the size of the ridge for a fixed
plate thickness.  Other scaling laws such as the dependence of the
ridge width and energy on the dihedral angle were also investigated.

In this article we extend the analytical and numerical study of the
ridges to explore whether the results obtained for isolated ridges can be
successfully applied to a network of interacting ridges in a crumpled
elastic sheet.  In Sec.\ II we introduce the concept of the
``minimal ridge'' that refers to the
necessary and sufficient conditions leading to the formation
of the ridge singularity in the limit of the vanishing plate
thickness.  Deviations from these ``minimal'' boundary conditions
in a crumpled sheet can then be treated, at least in the first
approximation, as perturbations to the ``minimal'' ridge.
In Sec.\ III we discuss a ``virial theorem'' for ridges
that is a direct consequence of the energy scaling argument.  It provides
a useful tool in testing the confinement of elastic energy and the degree
to which ridge interaction influences it.
In Sec.\ IV we present a scaling argument and an extension of the
asymptotic analysis of the \vonK equations
that establish that there is a long range
decay of the longitudinal ridge curvature and the transverse stress far away
from the ridge.  The energy in these ridge ``echoes'' is negligible
compared to the energy of the ridge.  A perturbation framework for
dealing with external forcing of the minimal ridge
is presented in Sec.\ V.  A particular case of the ridge
compression is studied in detail and the scaling of the
elastic energy correction is found.
We show in Sec.\ VI that the small echo strains and
curvatures are sufficient to change
the energy of a far away ridge by a finite fraction.
Numeric evidence corroborating some of the claims made in the preceeding
chapters is presented in Sec.\ VII.
Finally, implications of the ridge properties for the crumpling problem
as well as future work are discussed in Sec.\ VIII.

\section{The ``minimal'' ridge}

To facilitate a study of the ridge singularity one must first write
down equations that describe large deflections of thin plates.
Second, an understanding of the conditions that lead to the formation
of the ridge singularity must be achieved.  It was suggested
in Ref.\ \cite{vonK} that the existence of the
sharp vertices where both radii of curvature are of the order
of the sheet thickness is a necessary and sufficient condition
for the formation of ridges.  Ridges connect these points of high
curvature.  A minimal way to create a ridge,
therefore, Ref.\ \cite{vonK} argued, is to introduce
sharp points at the boundary of a flat rectangular piece of elastic
material by requiring that its boundary follow a frame that
has a sharp bend. The \vonK equations that describe large deflections
of thin elastic plates can be used to deduce the asymptotic behavior
of the ridge solution in this simple geometry.

Let us recall the boundary value problem that exhibits the
ridge singularity \cite{vonK}.  Consider a strip
made of isotropic homogeneous elastic material with Young's modulus
$Y$ and Poisson ratio $\nu$.  It has thickness $h$ and
width $X$.  The points are labeled by the material coordinates
$(x,y)$ so that $x \in (-X/2, X/2)$.  The strip extends in the 
$y$-direction.  Normal forces are applied to the long boundaries
$x = \pm X/2$ in such a way as to force the boundary to follow a
rectilinear frame that has a sharp bend at $y=0$.  The bend dihedral
angle is $\pi -2\alpha$ (so that $\alpha=0$ corresponds to a flat
strip).
The shape of the strip and the elastic stresses are found
from a solution to the non-dimensionalized \vonK equations
\begin{mathletters}
\label{vonKeqs}
\begin{equation}
\nabla^4 f = [\chi,f]
\end{equation}
\begin{equation}
\lambda^2 \nabla^4 \chi = -{1 \over 2}[f,f].
\end{equation}
\end{mathletters}

\noindent  Here $\nabla^4 \equiv \nabla^2\nabla^2$ and 
a square bracket $[a,b]$ denotes a symmetric contraction
of the second derivatives of the fields $a$ and $b$

\be
[a,b]  & \equiv &  \epsilon_{\alpha\mu}\epsilon_{\beta\nu}
(\partial_\alpha\partial_\beta a)
(\partial_\mu\partial_\nu b) \nonumber\\
& = & {\partial^2 a \over \partial x^2} 
{\partial^2 b \over \partial y^2} +
{\partial^2 a \over \partial y^2}
{\partial^2 b \over \partial x^2} -
2{\partial^2 a \over \partial x \partial y}
{\partial^2 b \over \partial x \partial y}.
\ee

\noindent
Here $\epsilon_{\alpha\beta}$ is the antisymmetric two by two tensor.
Summation over repeated indices is implied.

All lengths are measured in terms of the strip width $X$ and energies
in terms of the bending rigidity $\kappa = Yh^3/(12(1-\nu^2))$.
The first \vonK equation is the statement of the local
normal force equilibrium.  The second one has a purely
geometric origin.  It simply states that Gaussian curvature
$-{1 \over 2}[f,f]$ acts as a source for the stress field.
The small parameter $\lambda \sim h/X$ is proportional to the dimensionless
thickness of the sheet.
Here $f(x,y)$ and $\chi(x,y)$ are the potentials whose derivatives
give the curvatures $C_{\alpha\beta}$ and the two dimensional in-plane
stresses $\sigma_{\alpha\beta}$ via

\begin{equation}
C_{\alpha\beta} =  X^{-2}\ \partial_\alpha \partial_\beta f
\end{equation}
\begin{equation}
\sigma_{\alpha\beta} = \kappa X^{-2}\ \epsilon_{\alpha\mu}\epsilon_{\beta\nu}
\partial_\mu\partial_\nu \chi.
\end{equation}

\noindent To clarify the meaning of the curvature tensor $C_{\alpha\beta}$
we note that its eigenvalues are the inverses of the principal radii of
curvature of the sheet.  The sheets assumes a conformation that minimizes the
elastic energy consisting of the bending and the stretching parts (measured
in the units of $\kappa$)

\begin{mathletters}
\label{energies}
\begin{equation}
E_{\rm bend} = \int dxdy [\nabla^2 f]^2
\end{equation}
\begin{equation}
E_{\rm str} = \lambda^2 \int dxdy [\nabla^2 \chi]^2.
\end{equation}
\end{mathletters}

There are two ways to supply boundary conditions
for the \vonK plate equations consistently.  First, one could specify
a Kirchoff type condition on the functions $f$ and $\chi$ and their
derivatives in terms of the {\it material} coordinates.  Second,
a prescribed shape in the {\it embedding} space corresponds
to clamping the boundaries of the strip or simply supporting
them.  Only the first type of the boundary conditions is
tractable in general since the relation between the derivatives
of the functions $f$ and $\chi$ and the shape of the sheet in 
the embedding space is non-linear.  In addition, even if one 
succeeds in translating the boundary conditions that make reference
to the embedding space into the language of $f$ and $\chi$,
they will {\it change} when the thickness of the sheet is
varied or external forces are applied.
Therefore, only when the stresses and curvatures are
{\it specified} at the boundary can one make any
analytical progress.
We must remark at this point that the effective boundary
conditions for a ridge in a crumpled sheet are not of this
type.  Stresses in the facets are non-zero.  More importantly,
they depend on how each ridge is stressed by the rest of the
sheet.

The ``minimal ridge'' boundary conditions are given in terms of
the curvature at the boundary.  The only requirement is that there
be two sharp points on each long boundary of the strip.
The stresses and torques vanish at the boundary $x=\pm X/2$
(except at the singular point $y=0$)

\begin{equation}
\label{BC-minimal}
\partial_\alpha\partial_\beta\chi = \partial_\alpha\partial_\beta f = 0.
\end{equation}

\noindent 
The sharp vertices introduced at the boundary can be mathematically
expressed as singularity in the curvature at the boundary points
$y=0$ and $x=\pm X/2$

\begin{equation}
\label{BC-singular}
{\partial^2 f \over \partial y^2} = \alpha \delta(y).
\end{equation}

\noindent  The coefficient $\alpha$ in Eq.\ (\ref{BC-singular})
is exactly the bending angle alpha of the frame which is equal to the
half of the difference of the dihedral angle of the frame from $\pi$
\cite{vonK}.

The motivation behind seeking the ``minimal ridge,''
is that a number of the asymptotic properties
of the ridge singularity are independent of the details of the
boundary conditions.  These include the exponents in the
asymptotic thickness scaling of the ridge curvature, elastic
energy and other quantities.  Dihedral angle scaling exponents
are also independent of the details of the boundary conditions
\cite{vonK}.  Other ridge properties,
that do depend on the details of the boundary conditions
(the longitudinal stress supported by a ridge, for example),
can then be found perturbatively, at least in the first approximation.
The following two sections explore additional boundary
conditions-independent asymptotic properties of the ``minimal ridge''
that are useful in the investigation of the effects of the crumpled
sheet environment on the ``minimal ridge.''

\section{Virial theorem}

A study of the ``minimal ridge'' \cite{vonK} revealed that
as the thickness of the sheet vanished, the shape developed a sharp
crease.  Details of how that singular limit is approached were found.
In particular, the elastic energy concentrates
in a small region around the ridge of the size 
given by the characteristic ridge curvature $C$.  This fact,
allows one to construct an energy scaling argument that yields
the asymptotic scaling behavior
of the ridge curvature $C$ and the elastic energy $E$
\cite{fullerine}.

For the purpose of the energy scaling argument given in Ref.\
\cite{fullerine}, let us rewrite the expressions for the bending
and the stretching energies Eqs.\ (\ref{energies}) in terms of the
principal strains $\gamma_1$ and $\gamma_2$ and the
principal radii of curvature $R_1 \equiv 1/C_1$ and $R_2 \equiv 1/C_2$.
We get \cite{landau}

\begin{mathletters}
\label{energies-scale}
\begin{equation}
E_{\rm bend} = {1 \over 2}\int dS\ 
\left(\kappa (C_1 + C_2)^2 + \kappa_G C_1C_2\right)
\end{equation}
\begin{equation}
E_{\rm str} = {1 \over 2}\int dS\ 
\left(G(\gamma_1 +\gamma_2)^2 +G_s\gamma_1\gamma_2\right).
\end{equation}
\end{mathletters}

\noindent $G = Yh$ is the two dimensional stretching modulus
of a sheet of thickness $h$ made of elastic material with 
Young's modulus $Y$.
For the purposes of an energy scaling argument we may ignore the
second terms in the expressions for the energies. 
The argument given by Witten and Li in Ref.\ \cite{fullerine}
estimates these energies in terms of a characteristic
curvature $C$, for example, the transverse curvature in the
middle of the ridge.  According to that argument,
if the length of the ridge, {\it i.e.}\ the distance between the vertices,
is $X$, then the characteristic strain  $\gamma \sim (CX)^{-2}$ exists
in the ridge region of width $C^{-1}$ so that
the total stretching energy in the ridge is approximately
$E_{\rm str} \sim G \gamma^2 (X/C) \sim G X^{-3}C^{-5}.$
Similarly, the bending energy is given by the characteristic ridge 
curvature $C$, via $E_{\rm bend} \sim \kappa C^2 (X/C) \sim 
\kappa XC.$  Ref.\ \cite{fullerine} then argued that $C$ is the
only parameter that characterizes the ridge.  Also, since the
bending and the stretching energies both vary as a power of $C$, 
they must be comparable when the total energy is minimized.  
It immediately followed that $C \sim (\kappa/G)^{-1/3}X^{-2/3}
\sim h^{-1/3}X^{-2/3}$. 

Another important conclusion of the energy scaling
argument emerges when we consider the derivative of the energy with 
respect to the parameter $C$.  It must vanish for the value of $C$
at which the minimum total energy is achieved.  Since the energies
depend on powers of $C$ we obtain the following statement

\begin{equation}
{dE \over dC} = {dE_{\rm str} \over dC} + {dE_{\rm bend} \over dC} = 
{-5E_{\rm str} + E_{\rm bend} \over C}= 0,
\end{equation}

\noindent
which leads to a ``virial theorem''  for ridges $E_{\rm bend} = 
5 E_{\rm str}$ analogous to the virial theorem relating kinetic to
potential energy in celestial mechanics \cite{landau-mechanics}.
The validity of this virial theorem rests on rather general grounds.
The only requirements are that a) the energy be rigorously
expressible a sum of a bending and a stretching contributions
and b) each of these varies as a {\it power} of a free parameter, here $C$.
Thus a failure of the virial theorem would indicate that the
energies did not vary with the anticipated powers of $C$.
Such a failure would be expected if $C^{-1}$ were comparable
to other lengths in the problem, such as the size of the sheet $X$.

We have performed a numerical test of the virial theorem using
a lattice model of an elastic sheet after Seung and Nelson \cite{seungium}.
A numerical verification of the virial theorem for the
tetrahedral shape is presented in Sec.\ VI, here we only remark that
the agreement is better than few percent.
If a view of the crumpled membrane as a
network of ridges that meet at vertices is correct, the tetrahedron
virial theorem suggests that ridges in a crumpled sheet are well defined
objects to which a scaling argument can be applied.

\section{Large distance behavior of the ridge solution}

To ascertain how much ridges influence one another in
a crumpled sheet one must determine how the boundary layer solution,
that is valid in progressively narrow region around the ridge midline,
joins onto the flat solution far away from the ridge.
The following development is motivated by an observation
that in a numerical implementation of a ridge,
the ``echo'' strains and curvatures persist far away from the ridge, 
despite the fact that most of the energy is concentrated
into a region of asymptotically vanishing width.  In fact,
the characteristic decay length $L$ of the ``echo'' strains and curvatures
may diverge in the $\lambda \rightarrow 0$ limit.
In this section we present an energy scaling argument for the existence and
asymptotic scaling of the decay length $L$ for the ``minimal ridge''
configuration.  This conclusion is put on a more rigorous footing by
an extension of the asymptotic analysis of the
\vonK equations that includes the matching condition
between the boundary layer and the large distance solution.
The scaling of the long distance solution is later 
supported by numerical evidence.

We imagine cutting the ridge along its midline into two identical parts.
For the boundary value problem defined on one of resulting
semi-infinite strips, the ridge effectively introduces some
complicated boundary conditions applied at the cut.
The detailed form of these boundary conditions is not important for our
purposes.  It is significant, however, that they have no singularities.
For the purposes of an energy scaling argument, we imagine that
the short side of the semi-infinite strip is slightly bent so that
its middle is displaced in the normal direction by an amount $a$
as shown on Fig.\ 1.
\begin{figure}
\centerline{\epsfxsize=8cm \epsfbox{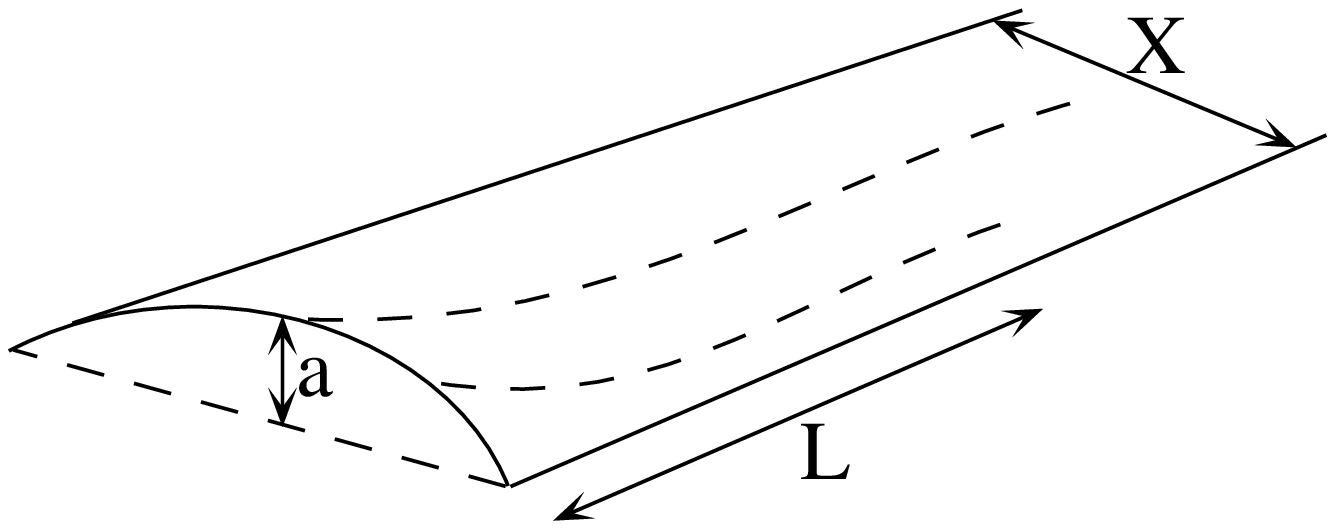}}

FIG.\ 1. Gentle curvature at a short end of a semi-infinite
strip decays at a characteristic distance $L$.
\end{figure}
This curvature decays to zero at a characteristic distance $L$
from the short end of the strip.  This length is set by the competition
of the bending energy that favors quick decay of the curvature
and the stretching energy that is smallest when the decay length is large.
The bending energy is given by the
integral of the squared mean curvature over the area of the strip.
If the decay length $L$ is much larger than the vertical displacement $a$,
the dominant curvature is in the $x$ (short) direction and is approximately
$C_{xx} \sim a/X^2$.  The bending energy is then
$E_{\rm bend} \sim \kappa C_{xx}^2 XL \sim \kappa a^2 X^{-3} L$.  
Here $\kappa$ is the bending rigidity and 
$XL$ is the area of the strip where the deformation exists.  The
strain created by the decay of the transverse curvature $C_{xx}$ is
due to the fact that the middle of the sheet is inclined by a small angle
of order $a/L$ and thus the projected length of the line that bisects
the strip is shorter than the length of the same segment
of the boundary.  The length mismatch creates a characteristic strain
in the $y$ (long) direction $\gamma_{yy} \sim a^2/L^2$.
Thus the stretching energy is $E_{\rm str} \sim G \gamma_{yy}^2 XL
\sim \kappa h^{-2} a^4 X L^{-3}$.  Here we used the fact that the
2D stretching modulus $G$ is related to the bending
rigidity via $G \sim \kappa/h^2$.
Since both kinds of energies vary as a power of $L$, they must
be comparable when the total energy $E_{\rm bend}+E_{\rm str}$
is minimized.  We thus, obtain the scaling of the
decay length $L \sim X (a/h)^{1/2}$.  The displacement $a$ is
determined by the scaling properties of the ridge.  The
asymptotic scaling of $C_{xx} \sim a/X^2$ must be consistent with that of the
the longitudinal ridge curvature
$\partial^2\! f/\partial x^2 \sim \lambda^{1/3}/X$.
Therefore $a \sim X\lambda^{1/3}$ which yields the scaling for the decay 
length $L$ and the elastic energy in the ridge ``wing'' 
in the $\lambda \rightarrow 0$ limit

\begin{equation}
L \sim X \lambda^{-1/3} \hspace{2em} E_{\rm wing} \sim \kappa \lambda^{1/3}.
\end{equation}

\noindent
Notice that the energy in these ridge ``wings'' or ``echoes''
is negligible compared to the ridge energy which scales as
$\kappa\lambda^{-1/3}$ \cite{science}.
This energy is spread over an increasingly large area.

The scaling of the ridge ``wings'' can be also obtained from
an extension of the asymptotic analysis
of the \vonK equations.  Ref.\ \cite{vonK}
determined the scaling of the boundary layer solution by
rescaling all variables by a power of $\lambda$ as in Eq.\ (\ref{rescale})
and then requiring that the highest derivative terms in Eqs.\ (\ref{vonKeqs})
be of the same order in $\lambda$.  The exponents of the $\lambda$-factors
that rescaled $f$ and $y$ were identical due to the
imposed boundary condition $f=\alpha\vert y\vert$ (this a particular
way to satisfy Eq.\ (\ref{BC-singular})).
Here we are seeking the scaling of the large distance 
solution that is pieced together with the ridge solution.
Hence, we can find such a solution separately for either
side of the ridge.  This allows for arbitrary rescaling
factors for $f$ and $y$ since the large distance behavior
of $f$ in the case of the ``wing'' is no longer required to be linear in $y$.
The matching condition on say $\partial^2\! f /\partial x^2$ for some
fixed $y$ and $\lambda \rightarrow 0$ requires that the ``wing''
solution for $f$ scale with $\lambda$ in the same way as the ridge solution
{\it i.e.}\ $f = \lambda^{1/3}\tilde f$, where $\tilde f$ is finite in the
$\lambda \rightarrow 0$ limit.
The same reasoning applies to the asymptotic scaling of $\chi$.
Thus the rescaling transformation that is needed to determine
the scaling of $y$ in the ``wing'' solution is

\begin{equation}
\label{new-scale}
f = \lambda^{1/3} \tilde f,\ 
\chi = \lambda^{-2/3} \tilde\chi,\ 
y = \lambda^\beta \tilde y,\
x = \lambda^0 \tilde x.
\end{equation}

\noindent The longitudinal direction $x$ is not affected by the
rescaling transformation, and the exponent $\beta$ must be negative
since a positive $\beta$ would reproduce the boundary layer scaling.
Plugging the scaling ansatz Eq.\ (\ref{new-scale}) into the
\vonK Eqs.\ (\ref{vonKeqs}) we obtain

\begin{mathletters}
\label{plug-ansatz}
\be
&\lambda^{1/3}&\left[
{\partial^4 \tilde f \over \partial \tilde x^4}+
2\lambda^{-2\beta} 
{\partial^4 \tilde f \over \partial \tilde x^2 \partial \tilde y^2}+
\lambda^{-4\beta}
{\partial^4 \tilde f \over \partial \tilde y^4}\right] =\nonumber\\
&&=\lambda^{-2/3+1/3-2\beta}[\tilde\chi,\tilde f]
\ee
\be
&\lambda^{2-2/3}&\left[
{\partial^4 \tilde\chi \over \partial \tilde x^4}+
2\lambda^{-2\beta}
{\partial^4 \tilde\chi \over \partial \tilde x^2 \partial \tilde y^2}+
\lambda^{-4\beta}
{\partial^4 \tilde\chi \over \partial \tilde y^4}\right] =\nonumber\\
&&=-{1\over2}\lambda^{2/3-2\beta}[\tilde f, \tilde f].
\ee
\end{mathletters}

\noindent
Balancing the dominant terms we obtain $\beta = -1/3$ in contrast
to $\beta = 1/3$ for the boundary layer solution \cite{vonK}.
Therefore, the decay length $L$ of the wing strains and curvatures
scales as $X\lambda^{-1/3}$ in agreement with the prediction of the
energy scaling argument above.
The leading order behavior of the
elastic energy in the wings can be found by substituting the rescaled
variables into Eqs.\ (\ref{energies}).   We obtain
$E_{\rm wing} \sim \kappa\lambda^{1/3}$ in accord with the
energy scaling argument.
In Sec.\ VI we numerically verify the existence of the
long range decay of the curvatures and stresses away
from the ridge.  The decay length is found to
scale as predicted in the small thickness limit.

\section{Ridge under external forcing}

To assess the relevancy of the results obtained for the
``minimal ridge'' to ridges in a crumpled membrane, we must 
first discuss ways in which the effective boundary conditions
for a ridge in a crumpled sheet differ form that of the ``minimal
ridge,'' and second, we ought to determine how these differences
affect such relevant ridge properties 
as the coefficients in the thickness scaling laws.
The additions and changes to the rectilinear frame boundary conditions
for the ``minimal ridge,'' that distinguish them from realistic
boundary conditions for a ridge in a crumpled sheet,
can consist of a) stresses applied at the
boundary, b) torques applied at the boundary, and c) distributed normal
forces that arise when distant parts of the crumpled sheet
press on the ridge.

In this section we discuss the linear response of the ``minimal ridge''
to external perturbations.  {\it A priori}, there is no reason to believe
that the asymptotic scaling of the linear response moduli is
independent of the details of the boundary conditions that
create and, more importantly, maintain the ridge in the process 
of external loading.  In other words, if we have chosen to maintain
constant normal forces on the boundaries instead of maintaining
constant curvature, the scaling of the linear response moduli
with the thickness could be different.
In fact, numerical evidence suggests that the 
scaling of the ridge stiffness with respect
to compression does indeed depend on the way boundary
conditions are maintained as the ridge is being distorted.
One must therefore address the applicability of the results derived in
this section to determination of the elastic response of 
crumpled sheets where the details of the boundary conditions are not known.
We are unable to address this question here beyond suggesting that the
linear response of a regular tetrahedron considered in Sec.\ VII
might reflect the situation in the crumpled sheet given
our view of the crumpled membrane as a collection of vertices, ridges and
facets.

\subsection{Linear response to specified boundary forces}

We present a treatment that allows one to construct 
a consistent perturbation expansion in the small external
forces that act at the boundary.  The case of the distributed
external forces can treated in a similar
manner that will be discussed below.
This method of treating perturbations
is by no means unique.  It allows one, however, to easily establish
asymptotic scaling of the energy correction in the limit
of the vanishing membrane thickness.
The essential idea is to reformulate
the problem in terms of some new functions $f_i$ and $\chi_i$
that are subject the {\it unchanged} ``minimal ridge''
boundary conditions $B_0$ but satisfy modified equations.
Consider the case when the full boundary forces $B_{\rm ext}$
(including the applied external forces)
can be decomposed into $B_{\rm ext} = B_0 + \delta B$.
Let $f_e$ and $\chi_e$ be the solution to the boundary value problem 
$\delta B$ that
includes {\it only} the small additional forces (i.e. the strip is not bent).
We then seek the solution to Eqs.\ (\ref{vonKeqs}) subject
to the full boundary conditions in the form

\be
f = f_i +f_e \hspace{2em} \chi = \chi_i + \chi_e,
\ee

\noindent so that $f_i$ and $\chi_i$ are subject to
the same boundary conditions $B_0$ as the undisturbed ridge solution
but satisfy modified equations

\begin{mathletters}
\label{vonK-mod}
\begin{equation}
\nabla^4 f_i = [\chi_i,f_i] +[\chi_e,f_i] + [\chi_i,f_e]
\end{equation}
\begin{equation}
\lambda^2 \nabla^4 \chi_i = -{1 \over 2}[f_i,f_i] - [f_e,f_i].
\end{equation}
\end{mathletters}

\noindent For a sufficiently small perturbation
one can construct a series expansion around the
unperturbed solution.

The equations for the first order corrections to the ridge
solution are linear and inhomogeneous.  The coefficients
as well as the source terms are proportional to the
second derivatives of the unperturbed ``minimal ridge''
solution.  These second derivatives are proportional to the
stresses and curvatures in the unperturbed ridge.
According to of Ref.\ \cite{vonK} the dominant
(dimensionless) stresses and curvatures are of
order $\lambda^{-1/3}$ in the region of the ridge of width
$\lambda^{1/3}$.  Sec.\ IV of this article determines
the behavior of the stresses and curvatures in the rest of
the sheet.  The dominant non-dimensional stresses and curvatures in the ridge
``wings'' are of order $\lambda^{1/3}.$   These ``echo'' disturbances
are spread over a large region of size $L \sim \lambda^{-1/3}$ so that
the elastic energy in the ``wings'' is negligible compared to the 
ridge energy.
The correction to the ridge solution will possess the same
qualitative features as the source terms in the equations (\ref{vonK-mod})
that determine them.  Since the behavior of the source terms in Eqs.\ 
(\ref{vonK-mod}) is completely determined by the ``minimal ridge'' solution
we postulate that the perturbing energy is confined to the region of the
ridge of width $\lambda^{1/3}.$
We will use this feature later to determine the appropriate
integration domain for the ridge energy correction
due to the external forces.

At this point we must mention how distributed forces
could be included in this treatment.  On the one hand, we could find
the effect of the external forces on the flat strip with 
free boundaries and then proceed with the derivation as above.
On the other hand, distributed normal forces can be directly
added to the first \vonK equation since it states the normal force balance
on an infinitesimal element of the sheet $dxdy$.
External in-plane forces lead to a redefinition of the Airy stress
function $\chi$.  It may not be at all possible to define a stress
function when in-plane external forces are present, however.
External distributed torques result in a redefinition of $f$
if its definition is indeed possible.  This is certainly the
case for small deflections when Monge coordinates can be used
\cite{vonK}.

The method presented above of splitting the boundary value
problem into two is applicable when several conditions
are satisfied.  First, note that if the equations
were linear, the result of such splitting would be trivial since
the equations for $f_i$ and $\chi_i$ would be the same as the
original equations and the result of the splitting the problem into
two amounts to the principle of superposition.
Second, the perturbed boundary
conditions $B_{\rm ext}$
that include the external force must be {\it compatible}
with the boundary conditions $B_0$.  For compatibility we require that
both sets of boundary conditions are supplied in terms of the same
functions (and derivatives) of $f$ and $\chi$ at the boundary.
Thus the change from $B_0$ to $B_{\rm ext}$ amounts to a change
in the boundary {\it values} of the specified functions.  We denote
the change in these boundary values schematically as 
$\delta B \equiv B_{ext} - B_0$.
This condition assures that the boundary value problem
$\delta B$ for $f_e$ and $\chi_e$ is well-posed.
Third, we assumed that the part of the boundary conditions $B_0$ that
create the ridge is unchanged in the process of loading.

It is unknown to us whether these conditions are satisfied
in a crumpled sheet.  We will see below that the prediction
based on this method, for the linear response to
point forces acting at the vertices
of a regular tetrahedron, is incorrect.
A possible explanation is that precisely the condition
for $B_0$ to be held constant in the process of loading
is violated in the tetrahedron.  The effective boundary
conditions for each ridge in a tetrahedron, the stresses
in the facets, for example, may change in proportion to the
forces acting at the vertices.
However, it might by possible to 
modify the conclusions of this method for a general
case of changing boundary conditions.  Certain features
that emerge from the perturbation scheme are likely
to survive for a general loading.  In particular,
the effect of the perturbation is likely to be confined
to the ridge region as above.

Another limitation of this method stems from the arbitrariness of the
choice of the external load.  In other words, of all possible ways
to load the ridge, the one in the ridge's ``weakest direction'' is
relevant to determination of the elastic response of
crumpled sheets.  It may happen that due to high symmetry, the
chosen load does not ``have a component in the
weakest direction'' of the ridge.  As a result, the stiffness
of the ridge in response to such a load is qualitatively greater
than its stiffness with respect to a generic load.
The concepts in quotes above can be made rigorous.
We discuss various types of loading in a later subsection with
regard to their relevance to the determination of the weakest ridge modulus.

\subsection{Point forces applied at the ridge vertices}

\begin{figure}
\centerline{\epsfxsize=8cm \epsfbox{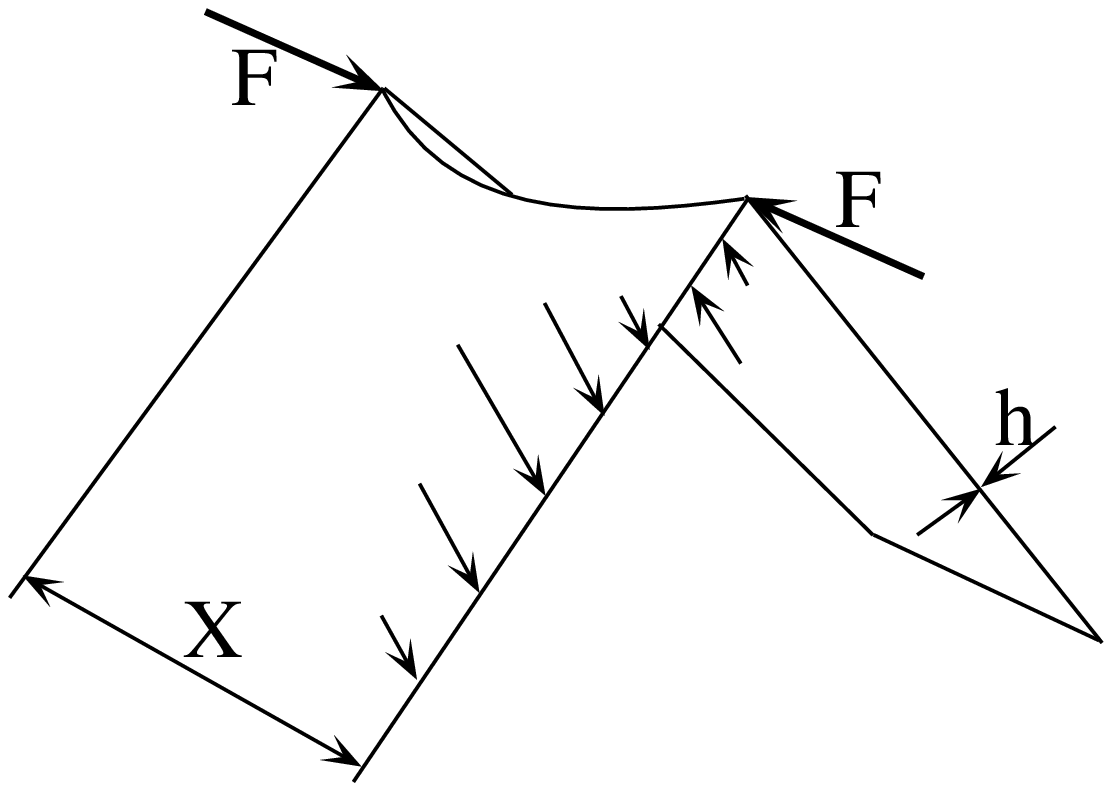}}

FIG.\ 2. A long strip of width $X$ is bent by normal boundary
forces (one quarter of which are shown) and is compressed
by forces $F$ applied at the vertices.
\end{figure}

Analysis can be carried further when the external influence
is characterized by a single small dimensionless parameter
$\bar F$.  In particular, for the most part
of the following development, we explicitly consider 
the case of compression of the ridge by a pair of point
forces acting at the ridge vertices as shown on Fig.\ 2.
Presumably, there are loads for which the
asymptotic $\lambda$-scaling of the solution to the ``flat'' problem
$f_e$ and $\chi_e$ is different from that in the case of the point forces.
This fact however introduces but a few changes in the following derivation.

A flat strip compressed by a pair
of point forces acting at the boundary remains
flat for small enough forces.  In addition, the stress function
is linear in the applied force $\bar F$

\begin{equation}
\label{f-and-chi_e}
f_e(x,y) = 0 \hspace{2em} 
\chi_e(x,y) = \bar F \lambda^{-1} \phi(x,y),
\end{equation}

\noindent where the function $\phi(x,y)$ depends neither on the
force $\bar F$ nor on the dimensionless thickness $\lambda$.
Above a buckling threshold for the compressing forces,
the flat solution Eq.\ (\ref{f-and-chi_e}) becomes unstable
to small perturbations.  Another buckled solution exists
and may have a different dependence on the force $\bar F$.
This fact does not affect the linear response analysis, however,
since the flat solution is unique for small $\bar F$.

We proceed with the boundary layer analysis following
Ref.\ \cite{vonK}.  This involves rescaling all variables
by a power of $\lambda$

\begin{equation}
\label{rescale}
\tilde f_i = \lambda^\beta f_i, \hspace{1em} 
\tilde\chi_i = \lambda^\delta \chi_i, \hspace{1em}
\tilde x = \lambda^0 x, \hspace{1em}
\tilde y = \lambda^\beta y,
\end{equation}

\noindent with $\delta = 2/3$ and $\beta = -1/3$ chosen in such a way
as to balance the powers of $\lambda$ in front of the terms
in the Eqs.\ (\ref{vonK-mod}) that are relevant in the boundary layer.
As in Ref.\ \cite{vonK},
we seek the solution to the rescaled equations
as a series expansion in $\lambda^{2/3}$
\begin{mathletters}
\label{lambda-expansion}
\begin{equation}
\tilde f_i = f_i^{(0)} + \lambda^{2/3} f_i^{(1)} + 
\lambda^{4/3} f_i^{(2)} + \ldots
\end{equation}
\vspace{-0.25in}
\begin{equation}
\tilde\chi_i = \chi_i^{(0)}  + \lambda^{2/3} \chi_i^{(1)} + 
\lambda^{4/3} \chi_i^{(2)} + \ldots.
\end{equation}
\end{mathletters}

The equations for the zeroth order terms $f_i^{(0)}$ and
$\chi_i^{(0)}$ read 

\begin{mathletters}
\label{vonK-zero}
\begin{equation}
{\partial^4 f_i^{(0)} \over \partial\tilde y^4}
= [\chi_i^{(0)}, f_i^{(0)}] + \lambda^{1/3}\bar F[f_i^{(0)}, \phi]
\end{equation}
\begin{equation}
{\partial \chi_i^{(0)} \over \partial \tilde y^4}
=  -{1\over 2} [f_i^{(0)},f_i^{(0)}].
\end{equation}
\end{mathletters}

\noindent Let us examine the source term in detail.

\be
\label{additional}
[f_i^{(0)}, \phi] & = &{\partial^2 f_i^{(0)} \over \partial \tilde x^2}
{\partial^2 \phi (\tilde x, y)\over \partial y^2} +
\lambda^{-2/3} {\partial^2 f_i^{(0)} \over \partial \tilde y^2}
{\partial^2 \phi (\tilde x, y)\over \partial \tilde x^2} \nonumber \\
&-&2\lambda^{-1/3}{\partial^2 f_i^{(0)} \over \partial\tilde x\partial\tilde y}
{\partial^2 \phi (\tilde x, y)\over \partial \tilde x \partial y}.
\ee

\noindent 
The derivatives of $\phi$ are evaluated at $x=\tilde x$
and $y= \lambda^{1/3}\tilde y$.
Since the rescaled variables are finite in the 
$\lambda \rightarrow 0$ limit,
$\phi$'s behavior near $y=0$
determines the leading order behavior of its
derivatives in Eq.\ (\ref{additional}) in the
$\lambda \rightarrow 0$ limit.
The magnitude of the perturbation term is thus governed by
$\tilde F = \lambda^\rho\bar{F}$ where the value of the exponent $\rho$
is determined by the behavior of the derivatives of $\phi$
near the ridge.  For example, suppose all derivatives of $\phi$ are
finite at $y=0$.  This is indeed the case for the compression of a
strip by a pair of point forces \cite{landau-ibid}.  Then, the second term in
Eq.\ (\ref{additional}) dominates so that $\rho=-1/3$.
This behavior need not be generic, since the derivatives
of $\phi$ may vanish at $y=0$ by reasons of symmetry.
Then the second term in Eq.\ (\ref{additional}) may vanish
so that a higher order term in $\lambda$ dominates.
This results in more positive value of $\rho$.

The solution $f_i^{(0)}$ and $\chi_i^{(0)}$ 
to Eqs.\ (\ref{vonK-zero}) can be sought as a series
expansion in the small parameter $\tilde F$.
Thus, the zeroth order term in the $\lambda^{2/3}$-expansion of the 
solution to the \vonK equations in the presence
of the external forces acting on the ridge is given by
(up to quadratic terms in $\tilde F$)

\begin{mathletters}
\label{f-and-chi-corr}
\begin{equation}
f^{(0)} = f_i^{(0)} + f_e \simeq \lambda^{1/3}(f_0 + \tilde F f_1 + 
\tilde F^2 f_2)
\end{equation}
\vspace{-0.25in}
\begin{equation}
\chi^{(0)} = \chi_i^{(0)}
 + \chi_e \simeq \lambda^{-2/3}(\chi_0 + \tilde F\chi_1 +
\tilde F^2 \chi_2) + \lambda^{-1} \bar F\phi.
\end{equation}
\end{mathletters}

\noindent The terms linear in $\tilde F$ completely characterize the linear
response of the ridge to this particular type of loading.  In general,
according to the postulated confinement property of the perturbing energy,
second derivatives of $f_1$ and $\chi_1$ possess the same qualitative 
features as those of $f_0$ and $\chi_0$.  In particular,
the dominant second derivatives of $f_1$ and $\chi_1$ are significant in
the ridge region of characteristic width $\lambda^{1/3}.$

Substituting the $\tilde F$-expansions into Eqs.\ (\ref{energies})
we obtain the expressions for the bending and stretching energies
accurate up to terms of order $\bar F^2$

\begin{mathletters}
\label{F-expansion}
\begin{equation}
E_{\rm bend} \simeq E_0^b + E_1^b \bar F + E_2^b \bar F^2
\end{equation}
\vspace{-0.25in}
\begin{equation}
E_{\rm str} \simeq E_0^s + E_1^s \bar F + E_2^s \bar F^2,
\end{equation}
\end{mathletters}

\noindent The energy of the unperturbed ridge as in Ref.\ \cite{vonK}
is $E_0^b \sim E_0^s \sim \lambda^{-1/3}$.
Other $E_i$'s are sums of integrals that involve
second derivatives of $f_i$, $\chi_i$ and $\phi$.  For example

\be
\label{correction}
E_2^s \bar F^2
& \simeq &\lambda^2 \int d\bar{x}d\bar{y}\ (\lambda^{-2/3}
\tilde F \nabla^2\chi_1 + \lambda^{-1} \bar F \nabla^2 \phi)^2 \nonumber \\
& = &\bar F^2 \int d\bar{x}d\tilde{y}\ \displaystyle{\left[
\lambda^{-1/3+ 2\rho}
\left({\partial^2 \chi_1 \over \partial \tilde y^2}\right)^2 +
\lambda^{\rho} \left({\partial^2 \chi_1 \over \partial \tilde y^2}\right)
\nabla^2\phi\right]} \nonumber\\
& + &\bar F^2 \int d\bar{x}d\bar{y}\ (\nabla^2\phi)^2.
\ee

\noindent We must remark that the terms quadratic in $\bar F$ that involve
$f_2$ and $\chi_2$ cancel each other so that the linear in $\bar F$ correction
to the potentials $f$ and $\chi$ determines the quadratic correction to the
total energy.  This situation is common to all linear response problems.

The leading order behavior of $E_1$ and $E_2$ in the small thickness
limit can be found by assuming, as before, that the corrections to the ridge
solution $f_1$ and $\chi_1$ are confined to the ridge region.
The domain of integration in $\bar y$ in the integrals involving these ridge
corrections as in Eq.\ (\ref{correction}) is thus of order $\lambda^{1/3}$.
We can therefore obtain the asymptotic scaling of the coefficients
in Eqs.\ (\ref{F-expansion}).  The bending and the stretching pieces
scale with $\lambda$ in the same way.  The correction to
the total energy is $E_2 \bar F^2$

\begin{mathletters}
\label{E-corrections}
\begin{equation}
E_1^{(b,s)} \sim A_1^{(b,s)} \lambda^{-1/3 +\rho} + B_1^{(b,s)}
\end{equation}
\vspace{-0.25in}
\begin{equation}
E_2 =E_2^s+E_2^b\sim A_2 \lambda^{-1/3 + 2\rho} + B_2\lambda^\rho + C_2,
\end{equation}
\end{mathletters}

\noindent where $A_1^s$ etc. are arbitrary coefficients that
do not depend on $\lambda$ in the $\lambda \rightarrow 0$ limit.
If external stresses and curvatures given by the
derivatives of $\chi_e$ and $f_e$ respectively are zero on the
ridge line and do not increase sufficiently rapidly away from the
ridge, the exponent $\rho$ will be positive and large so that
the energy correction will be dominated by the last terms in the
expressions for $E_1$ and $E_2$.  These terms are identical
to the energy of a flat sheet acted on by the external forces.
If $\rho < 1/6$, however, the energy change is dominated
in the small thickness limit by the interaction of the
ridge with the external forces given by the corrections to the
ridge solution $f_1$, $\chi_1$.

Eqs.\ (\ref{E-corrections}) predict that both kinds of energy separately
depend linearly on $\bar F$.  We demonstrate
numerically in Sec.\ VII that upon application of point forces at the
ridge vertices, each energy correction does indeed depend linearly
on the applied force with the coefficient that scales with $\lambda$
as predicted.
The total energy can depend only on the square of the applied
force since the boundary stresses and torques vanish for the undisturbed 
ridge so that the work done by the external force is exactly
the change in the ridge's elastic energy.
This means that the linear terms in $\bar F$ in the expressions 
(\ref{E-corrections}) for $E_{\rm bend}$ and $E_{\rm str}$ must 
cancel each other, {\it i.e.}\ $E_1^s = -E_1^b$ to all orders
in the $\lambda^{2/3}$-expansion ($A_1^s = -A_1^b$, $B_1^s = -B_1^b$ etc.).
The numerics reported in Sec.\ VII convincingly show that it is
indeed the case. The $\lambda$-scaling of the ridge's elastic
constant $E_2$ given in the Eq.\ (\ref{E-corrections}b) has also
been shown to be consistent with the numerics.

When external forces distort the ridge, additional
longitudinal strain results.  We first define a
``vertex-to-vertex'' strain $\gamma_v$
as the amount by which the
vertices move closer together divided by $X$.  We compare this quantity
with the additional strain in the sheet that is the change in
$\gamma_{xx} \simeq (1/Y) \partial^2 \chi/\partial y^2$.  It
can be found from the first order in $\tilde F$ correction to $\chi$ 
Eq.\ (\ref{f-and-chi-corr}).
We can infer $\gamma_v$ by inspection of the energy-force relation,
since the work done by $\bar F$ is the change of the energy $E_2\bar F^2$.

\begin{equation}
\label{add-strain}
\gamma_v \sim \lambda E_2\bar F 
\sim (A_2 \lambda^{2/3 + 2\rho} + B_2 \lambda^{1+\rho} + C_2\lambda)\bar F
\equiv G^{-1}\bar F,
\end{equation}

\noindent whereas from Eq.\ (\ref{f-and-chi-corr}) we obtain the
change in the longitudinal strain in the sheet

\begin{equation}
\label{add-strainxx}
\gamma_{xx}^{\rm add} \sim \lambda^{2/3 +\rho}\bar F
\end{equation}

\noindent
Eq.\ (\ref{add-strainxx}) gives another way to
estimate the exponent $\rho$ numerically.

This way of looking at the problem motivates the following consistency
check of the perturbing energy confinement.  An additional way of
determining the $\lambda$-scaling of the total energy correction
via $E_2 \bar F^2 \sim G(\gamma_{xx}^{\rm add})^2 wX/\kappa \sim
\lambda^{-2/3+2\rho} \bar w$ must
reproduce the scaling of $E_2$ obtained by explicit integration
as in Eq.\ (\ref{E-corrections}).  Here $\bar w = w/X$ is the
characteristic width of the region in which the perturbing energy
is confined.  Calculated by this method, the width $\bar w \sim
\lambda^{1/3}$ is independent of the exponent $\rho$ and
scales with $\lambda$ in the same way as the ridge width.

\subsection{Buckling threshold}

An important property of the ridge solution that is inaccessible
by this simple-minded perturbation scheme is the force required
to {\it buckle} the ridge.  In principle, one must solve Eqs.\
(\ref{vonK-zero}) and then perform a linear stability analysis of the
solution to determine when the ridge will buckle.  This task is
intractable analytically due to the complexity of the equations.  The
asymptotic scaling of the buckling threshold may nevertheless be
anticipated using the following argument.
The undisturbed ridge solution $f_0$ and $\chi_0$ is linearly
stable against shape perturbations.  Therefore, changes in this
solution that destroy stability are likely to be of the
order of the solution itself.  To induce such changes,
the additional terms in the Eqs.\ (\ref{vonK-zero}) for the loaded ridge 
must be comparable in magnitude to the rest of the terms in the equations.
Hence, the parameter $\tilde F$ that controls the magnitude
of the additional terms in the Eqs.\ (\ref{vonK-zero})
has to be of order unity $\tilde F \sim 1$.
This means that in the case of ridge compressed by vertex forces,
the buckling threshold force scales as

\be
\bar F_{\rm crit} \sim \lambda^{-\rho}.
\ee

This argument has a serious flaw.  The scaling of the
``flat-problem'' solution $f_e$ and $\chi_e$ may be different
for large forces $\bar F$ since additional solutions to the
``flat-problem'' may appear.  This may result in a different
value of the exponent $\rho$ and thus a different scaling of the
buckling threshold.
The conjectured scaling of the critical load has two important implications.
First, the additional longitudinal sheet strain at the buckling
threshold given by the Eq.\ (\ref{add-strainxx})
is of order of the strain that existed in the ridge prior to compression.
Second, the ridge energy correction at the buckling threshold value 
of the external force has the same asymptotic thickness scaling as the energy
in the undisturbed ridge.
This conclusion is supported numerically in Sec.\ VII.
A corollary of this statement is that the applied load on a ridge
does not have a decisive effect on its energy.  The ridge buckles
when its energy changes by a finite fraction.  We will see in the following
subsection that this argument applies to other types of loading as well.
Therefore, any two unbuckled ridges in a crumpled sheet with comparable lengths
and dihedral angles should have comparable energies, even though they have
different loads.

\subsection{Other types of loading}

To describe elastic properties of the ridge in a way that is
relevant to determination of the structure of the crumpled sheet we
must investigate other types of loading.  
We first consider other loads that can be treated with the 
perturbation scheme developed above.  These include normal
forces acting on the ``minimal ridge.''  Having done that,
we anticipate that in a crumpled sheet the assumptions
that lead to the scaling laws for the energy corrections are violated.

Let us first use the perturbation scheme developed above to 
discuss other loads.  The goal is to determine the ``weakest modulus''
of the ridge alluded to above.  The perturbation treatment of this section
relates the linear response of the flat sheet to the linear response
of the ridge.  We therefore anticipate that any perturbation that
causes a large response in a flat sheet will also be relevant to 
the determination of the ``weakest'' ridge modulus.  Since the bending
rigidity $\kappa \sim Yh^3$ of an elastic sheet vanishes faster in the $h
\rightarrow 0$ limit than the stretching modulus $G \sim Yh$, any
perturbation that causes the sheet to bend will create a large response.
We therefore consider normal forces on a ridge presented on
Fig.\ 3.

\begin{figure}
\centerline{\epsfxsize=8cm \epsfbox{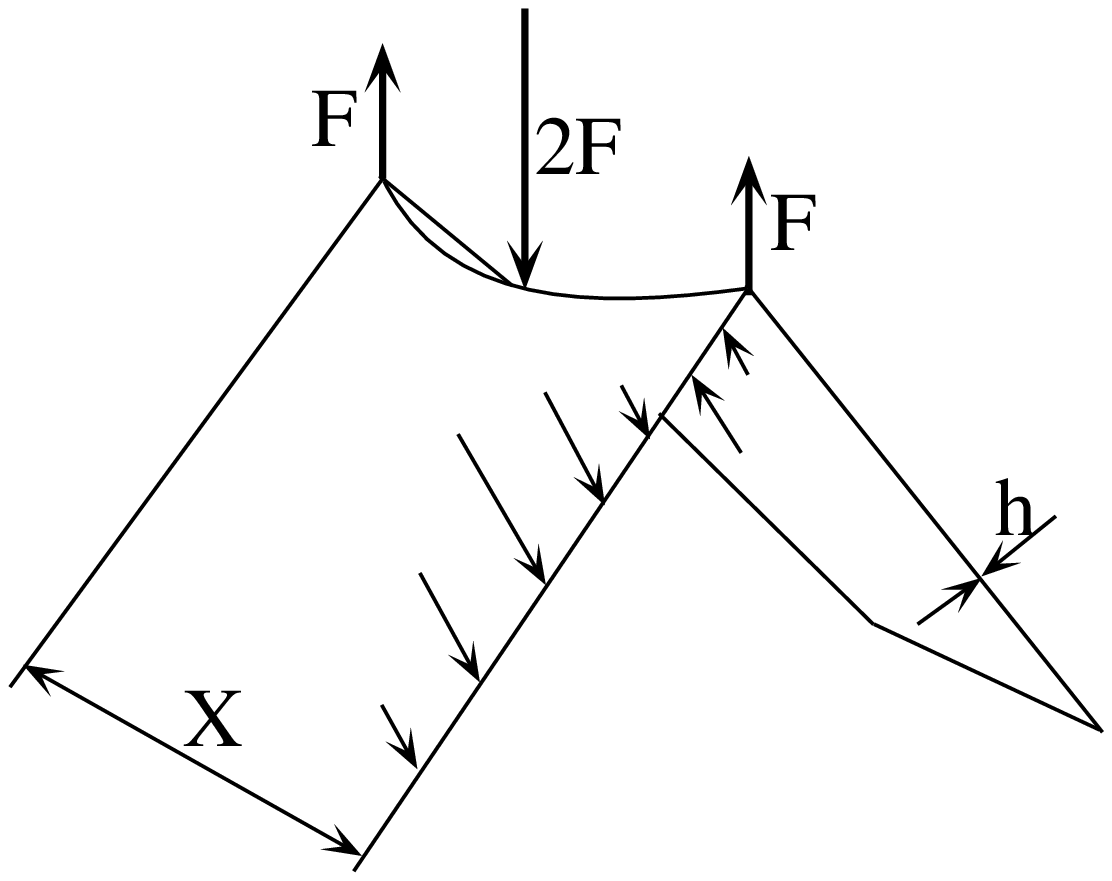}}

FIG.\ 3. Same strip is bent by normal forces applied at the
vertices as well as the middle of the ridge.
\end{figure}

Let us first describe the effect of these forces on a flat sheet.
The scaling of the curvature potential $f_e$ with force
$F$ can be found by noting that the torque of the
external force is balanced by the flexural moment of the sheet due
to its curvature $\kappa X\,\partial^2\! f_e/\partial x^2 \sim FX$.
Therefore, $f_e = \lambda^{-1} \bar F \varphi$ is controlled by a single
dimensionless parameter $\lambda^{-1}\bar F$.  Since the sheet is bent
in the short direction, we can use the arguments of Sec.\ IV to show
that this curvature persists up to a distance of order
$L \sim X\lambda^{-1/3}$.  The treatment of Sec.\ IV
 also determines the stresses in the strip.  The dominant component of
the stress is $\sigma_{yy} \sim Yh (a/L)^2$ where $a \sim f_e \sim
X\lambda^{-1}\bar F$.  Hence,
the stress is of the second order in the applied force $\bar F$.
The modified \vonK equations (\ref{vonK-mod})
will thus have a first order in $\bar F$ source
term due to $f_e$ and a second order in $\bar F$ source term due
to $\chi_e$.  One can now repeat the steps that lead from the modified
\vonK equations (\ref{vonK-mod}) via a rescaling transformation
Eq.\ (\ref{rescale}) to the perturbation expansion
Eqs.\ (\ref{F-expansion}).  The only difference is the form of the
flat solution $f_e$ and $\chi_e$.

Together with knowledge of the small $y$ behavior of the function $\varphi$ 
one can determine the exponent $\rho$ which controls the ridge stiffness
under this particular type of loading.  
The important second derivative $\partial^2\varphi/\partial x^2$ is
finite at the ridge.  Therefore, $\rho = -4/3$.
This means first that the linear response of the ridge to normal
forces is qualitatively greater than to in-plane forces.
Second, the normal force required to buckle
the ridge $\bar F_{crit} \sim \lambda^{4/3}$ is much smaller than
in the case of longitudinal compression.  The conjecture that the ridge
buckles when its energy changes by a finite fraction can be made on the
same basis as for the case of the longitudinal compression of the ridge. 

We have thus shown that depending on the symmetry of the applied load,
the linear response of the ridge can vary considerably.
It is unknown to us at this point whether the resistance of a 
crumpled sheet to further compression is determined solely by the
``weakest'' ridge modulus.  It is certainly not inconceivable to
imagine a situation in which a number of ridges in a crumpled sheet
that form a kind of a ``structural skeleton'' are loaded in such
a way that their linear response is given by a stronger modulus.

Let us finally discuss the applicability of the perturbation scheme
developed here to ridges in a crumpled sheet.
We anticipate the boundary stresses for the ridge in a
crumpled sheet to change by an amount proportional to a load.
We cannot, therefore, utilize the perturbation method, developed
in this section, to determine the scaling of the energy correction
and the buckling threshold.
Certain features of the conclusions that emerged from the
perturbation analysis are likely to persist, however.
We anticipate that, the perturbing energy is confined to the ridge region.
In addition, the change in the ridge longitudinal
strain (the dominant ridge strain) must
be proportional to the movement of the sheet caused
by the external forces.  The ``vertex-to-vertex'' strain
is an example of such movement.  The coefficient of
proportionality is likely to scale as a power of the thickness
$\lambda$.  In most situations, one can make a reasonable
guess to what that power is.  For example, the numerics reported in
Sec.\ VII show that when a tetrahedron
is compressed by point forces applied to its vertices,
the additional ridge strain scales in the same way with $\lambda$ as the
``vertex-to-vertex'' strain.  This assumption, together with
the notion of the perturbing energy confinement,
is sufficient to determine the scaling of the energy correction.

\section{Ridge Interaction}

Ridges in a crumpled elastic sheet can interact in two distinct ways.
First, distant parts of the sheet can press against a ridge.
This type of interaction can be discussed
using the framework developed in Sec.\ V.  However, a quantitative
understanding of the distribution of 
forces through the crumpled sheet is needed.
Second, nearest neighbor ridges can influence one another through
the strains and curvatures they create in the sheet.
We will concentrate on this particular aspect of the
ridge interaction since we are able to obtain quantitative
conclusions.

Two length scales characterize a ridge.  Most of the ridge's elastic energy
is confined within a strip of width $w \sim X\lambda^{1/3}$
around the ridge.  Small curvatures and strains persist up to a
distance $L \sim X\lambda^{-1/3}$ from the ridge.
In a crumpled elastic sheet typical distances between ridges $D$
are of the order of their length: $D\sim X$.  Assuming that
all ridges have the same characteristic size $X$ we are led to
a conclusion that $w \ll D \ll L$ for most ridges.
Therefore, the small residual strains and curvatures 
of a ridge will influence its neighbors in a way that is
calculable within the framework of Sec.\ V.
In fact, in Sec.\ IV we argued that there is a small transverse strain 
$\gamma_{yy} \sim a^2/L^2 \sim \lambda^{4/3}$ present in the
sheet.  This strain can be thought of as resulting from
an external force $\bar F \sim \lambda^{1/3}$
that is stretching the sheet in the direction
transverse to the ridge.  Therefore, the stress potential
function $\phi$ has a finite derivative
$\partial^2 \phi/\partial x^2$ at the ridge.
According to the prescription of Sec.\ V, $\rho = -1/3$,
so that the energy correction scales as 

\begin{equation}
\delta E \sim \kappa\lambda^{-1/3+2\rho}\bar F^2 \sim \kappa\lambda^{-1/3}.
\end{equation}

\noindent Note that the energy correction scales the same way with
$\lambda$ as the ridge energy.  This means that even though
the ridge ``wings'' carry a negligible amount of energy, they
cause the total energy of the system to be changed by a finite
fraction.  This is not surprising since the transverse ``echo'' stresses
discussed in Sec.\ IV are of the same magnitude as the transverse
stresses in the ridge.
The interaction energy of ridges in a strip geometry is studied numerically
in the following section.  It is found to be a few percent of the
ridge energy in the small thickness limit.

\section{Numerical simulation of the ridge properties}

In this section we use a lattice model of an elastic
sheet of Ref.\ \cite{seungium}.  This model was
used by the authors and others 
to verify a number of the ridge scaling properties \cite{science,vonK}.
The sheet is modeled by a triangular lattice of springs
of equilibrium length $b$ and spring constant $K$.
A bending energy $J(1 - {\bf \hat n}_1 \cdot {\bf \hat n}_2)$
is assigned to each pair of adjacent lattice triangles with normals
${\bf \hat n}_1$ and ${\bf \hat n}_2$.
When the strains are small compared to unity and radii of curvature
are large compared to the lattice spacing $b$, this model
bends and stretches like 
an elastic sheet of thickness $h = b \sqrt{8J/K}$
and bending modulus $\kappa = J\sqrt{3}/2$.
Ridges could be created either by imposing appropriate
boundary conditions to a long strip of the simulated material
or by introducing disclinations.  We studied both types of shapes.
First, we applied forces to the particles located on the
long boundaries of a strip so as to constrain them to lie in
different planes on each side of the ridge.  The angle between
the normals to these planes is $2\alpha$.
Second, we connected a triangular piece of this simulated material
into a regular tetrahedron so that each edge then became a ridge.
A sequence of minimum energy shapes of different dimensionless
thickness $\lambda$ were obtained
with the use of a conjugate gradient routine.
External forces were then applied to the ridge vertices
and linear response to compression was measured as well
as the buckling characteristics.

\begin{figure}
\centerline{\epsfxsize=8cm \epsfbox{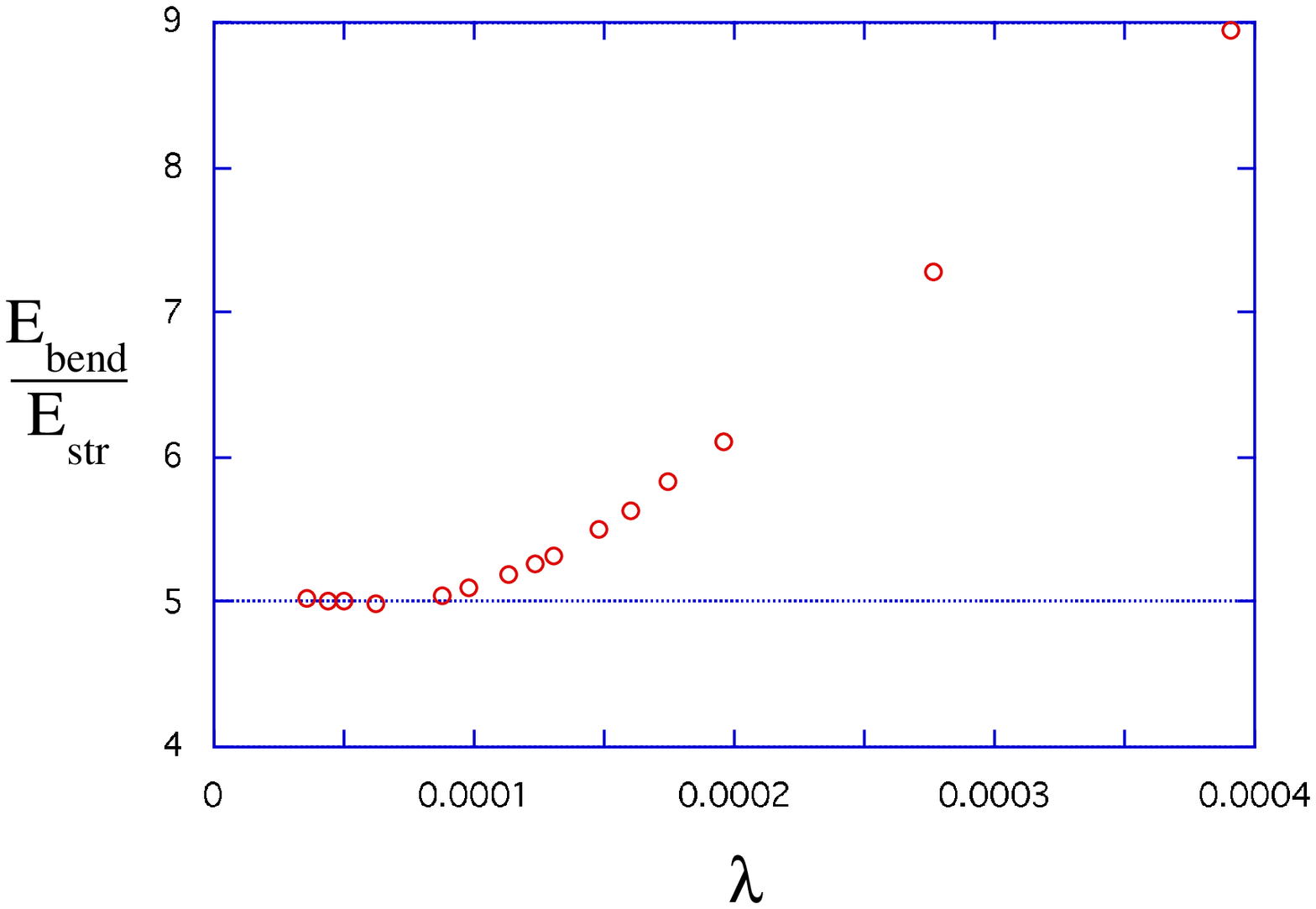}}

FIG.\ 4. Ratio of the total bending to the total stretching
energy vs.\ the dimensionless thickness $\lambda$ for a regular
tetrahedron of edge length $100b$.  The asymptotic limit of $5$
is approached for $\lambda \simeq 10^{-4}$.
\end{figure}

We first tested the virial theorem that predicts that
if all of the elastic energy is concentrated in the ridges
then the total bending energy is five times
the total bending energy.
In Fig.\ 4 we plot the ratio of the total bending to
total stretching energies for the tetrahedron with
edge length of $100b$.
This ratio approaches the predicted value of $E_{\rm bend}/E_{\rm str} = 5$
within a few percent for the values of dimensionless thickness
$\lambda \leq 10^{-4}$.  For the smallest values of $\lambda$ on the
plot the ratio deviates from the asymptotic value due to 
lattice effects.  Thus the virial theorem is obeyed within our
measurement precision.

The predictions of the scaling behavior of the ridge under
external loading was done using a $43.3b$ by 
$100b$ strip bent by boundary forces as well as
a tetrahedron with an edge length of $50b$.  
We applied point forces to the vertex particles in either case
in such a way as to compress the ridge.
We first compressed a flat sheet to verify that stresses 
are finite on the line $y=0$.  
Then, the prediction for the exponent $\rho$ defined in Sec.\ V is
$\rho=-1/3.$  The correction to the total elastic energy was found
to vary on $\bar F^2$ as anticipated.
The coefficient $E_2$ of $\bar F^2$ should scale with $\lambda$ according
to Eq.\ (\ref{E-corrections}).  Fig.\ 5 is a plot of $E_2$
as extracted by a quadratic fit of the correction of the
total elastic energy per ridge
for the tetrahedron (triangles) and the strip (squares)
as a function of $\lambda$.  The inset of Fig.\ 5
shows the fit to the quadratic dependence of the total energy
on the applied force.

\begin{figure}
\centerline{\epsfxsize=8cm \epsfbox{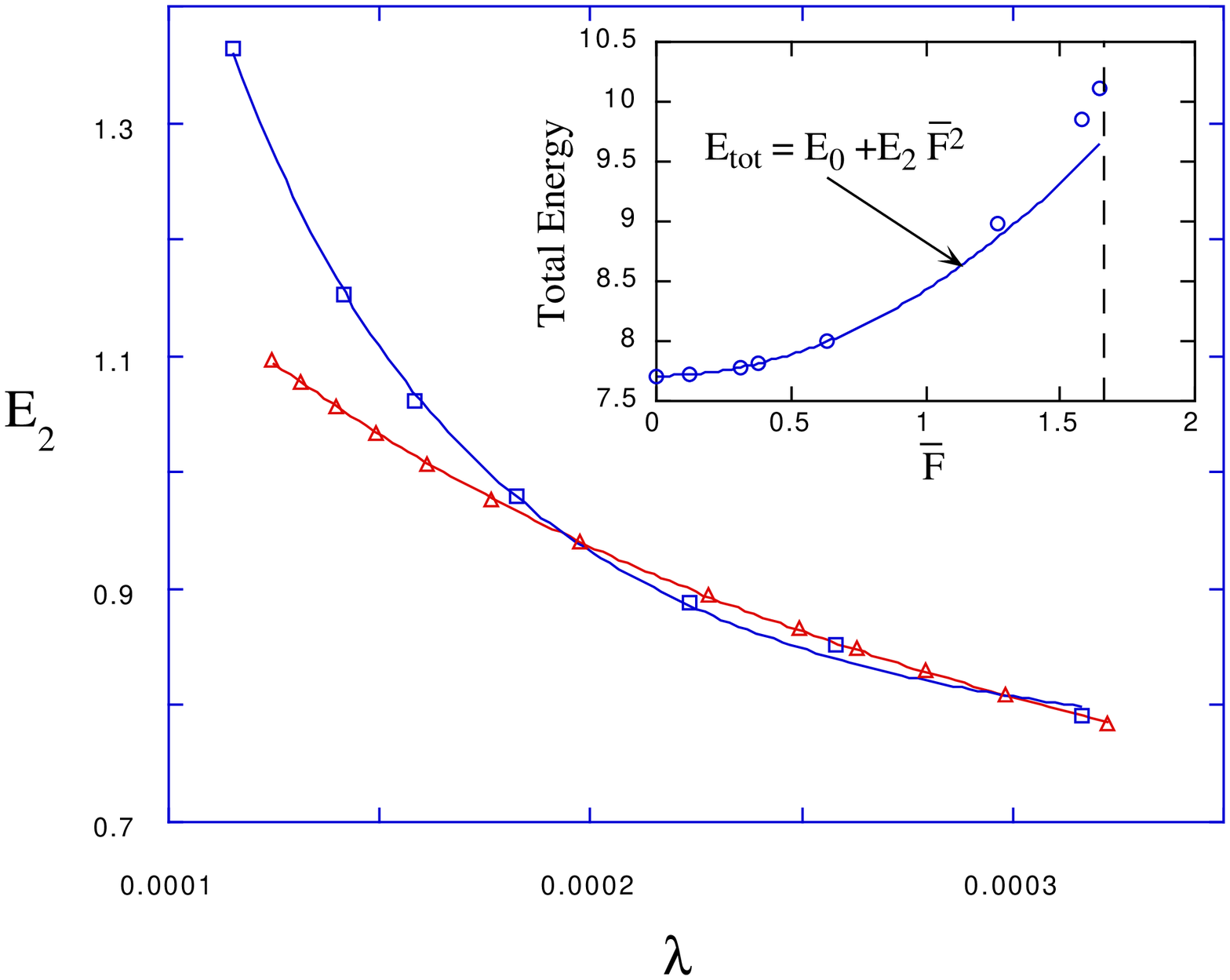}}

FIG.\ 5. Thickness dependence of the 
coefficient $E_2$ in $E/\kappa \sim E_0 + E_2\bar F^2$
for the strip of dimensions $43b$ by $100b$ (squares)
and tetrahedron of edge length $50b$ (triangles).  The solid
lines are fits according to the Eq.\ (\ref{E-corrections})
with $\rho=-1/3$ for the strip and $\rho=0$ for the tetrahedron.
Inset: Plot of the total energy of a strip as a function of the
applied force $\bar F.$  Solid line is the quadratic fit to the
first five points.  Dashed vertical line marks the buckling threshold.
\end{figure}

The numerically accessible range of $\lambda$ does not allow 
for a direct determination of the exponent $\rho$ from the data.
However, the strip data are consistent with the prediction
for $\rho = -1/3$.  However, the tetrahedron data are consistent 
with $\rho = 0$ and inconsistent with $\rho=-1/3$ when fitted with the
scaling form Eqs.\ (\ref{E-corrections}) and (\ref{add-strain}).
A given load appears to store qualitatively less energy for the
tetrahedron than for the strip, i.e. the tetrahedron is qualitatively
stiffer.  This difference might be explained by noticing that while
in the case of the strip, the boundary shape $B_0$ that maintained
the ridge was fixed under the loading, whereas the effective
boundary shape for the ridges in a tetrahedron changes when
forces are applied.  In addition, tangential stresses act on the
the effective ridge boundary in a tetrahedron, whereas only normal
boundary forces are present in the strip geometry.  Therefore, the
decomposition method of Sec.\ V cannot be used for the tetrahedron.
The scaling of the tetrahedron stiffness implies that the additional
strain due to the action of the compressive force is confined to the
ridge region and scales the same way with the thickness $\lambda$
as the ``vertex-to-vertex'' strain.
These results are different from the
situation in the strip.  There, the ``vertex-to-vertex'' strain
may be relaxed through bending so that the additional strain in the
sheet is qualitatively weaker than the movement of the vertices
would dictate if such relaxation was not possible.

\begin{figure}
\centerline{\epsfxsize=8cm \epsfbox{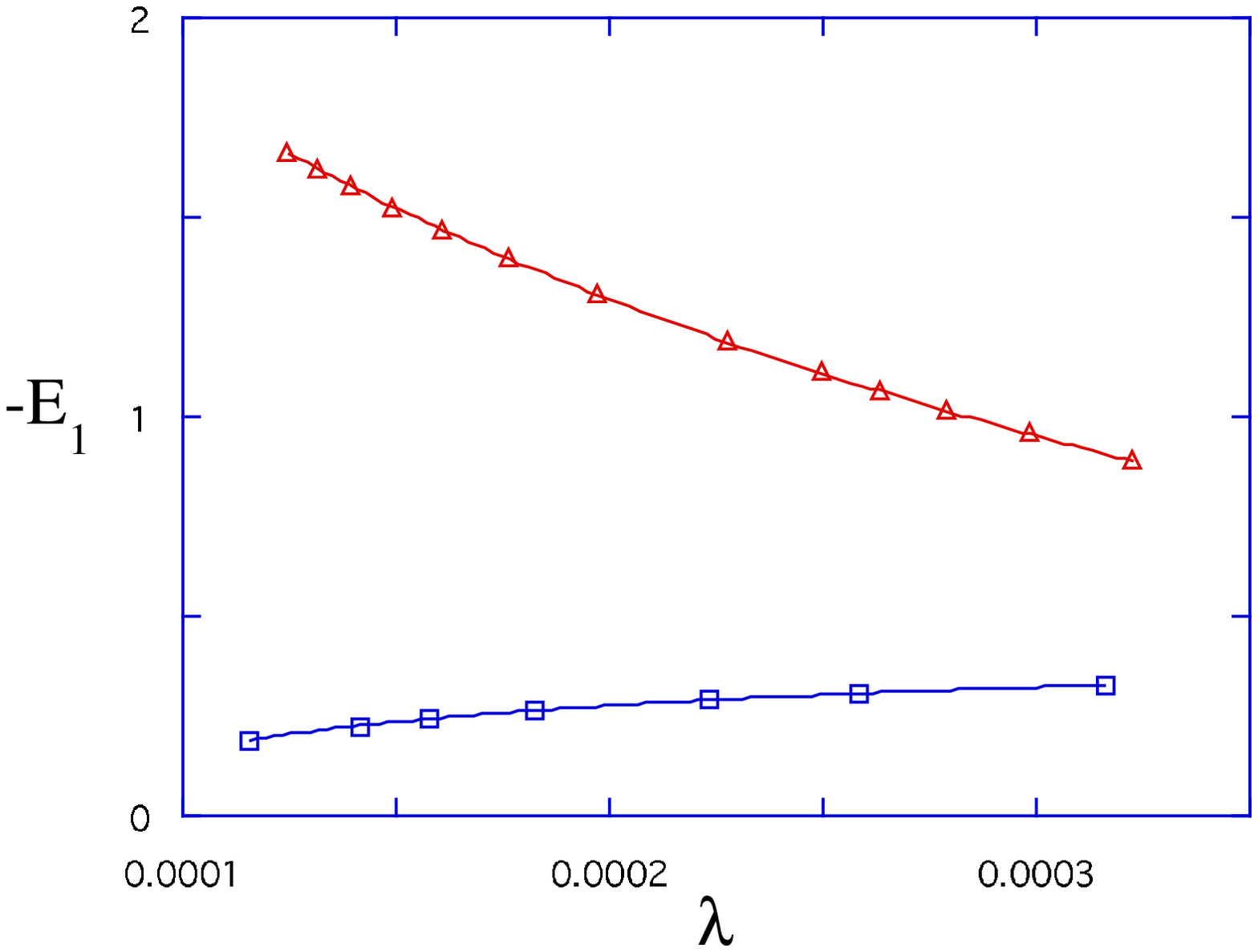}}

FIG.\ 6. The coefficient $E_1^b$ of the linear dependence of the
total bending energy on the force $\bar F$ for the $43b$ by $100b$
strip (squares) and the tetrahedron of size $50b$ (triangles).  The solid
lines are again fits according to the Eq.\ (\ref{E-corrections})
with the same values of the exponent $\rho$ as in Fig.\ 5.
\end{figure}

In Fig.\ 6 we plot $E_1^b$ for both shapes extracted
by a linear fit of the total bending energy as a function of $\bar F$.
The data are consistent with the prediction Eq.\ (\ref{E-corrections})
for the same values of the exponents $\rho$ as obtained from the scaling
of $E_2$.
Finally, we graph the coefficient $G^{-1}$ of the linear dependence of
the induced ``vertex-to-vertex'' strain on the applied force
in Fig.\ 7.
The data are consistent with the prediction Eq.\ (\ref{add-strain}).
For comparison let us consider a flat strip.
Its dimensionless inverse stiffness $G^{-1}$
to compression diverges as $\lambda^{-1}$, whereas
its stiffness to bending vanishes as $\lambda$.

\begin{figure}
\centerline{\epsfxsize=8cm \epsfbox{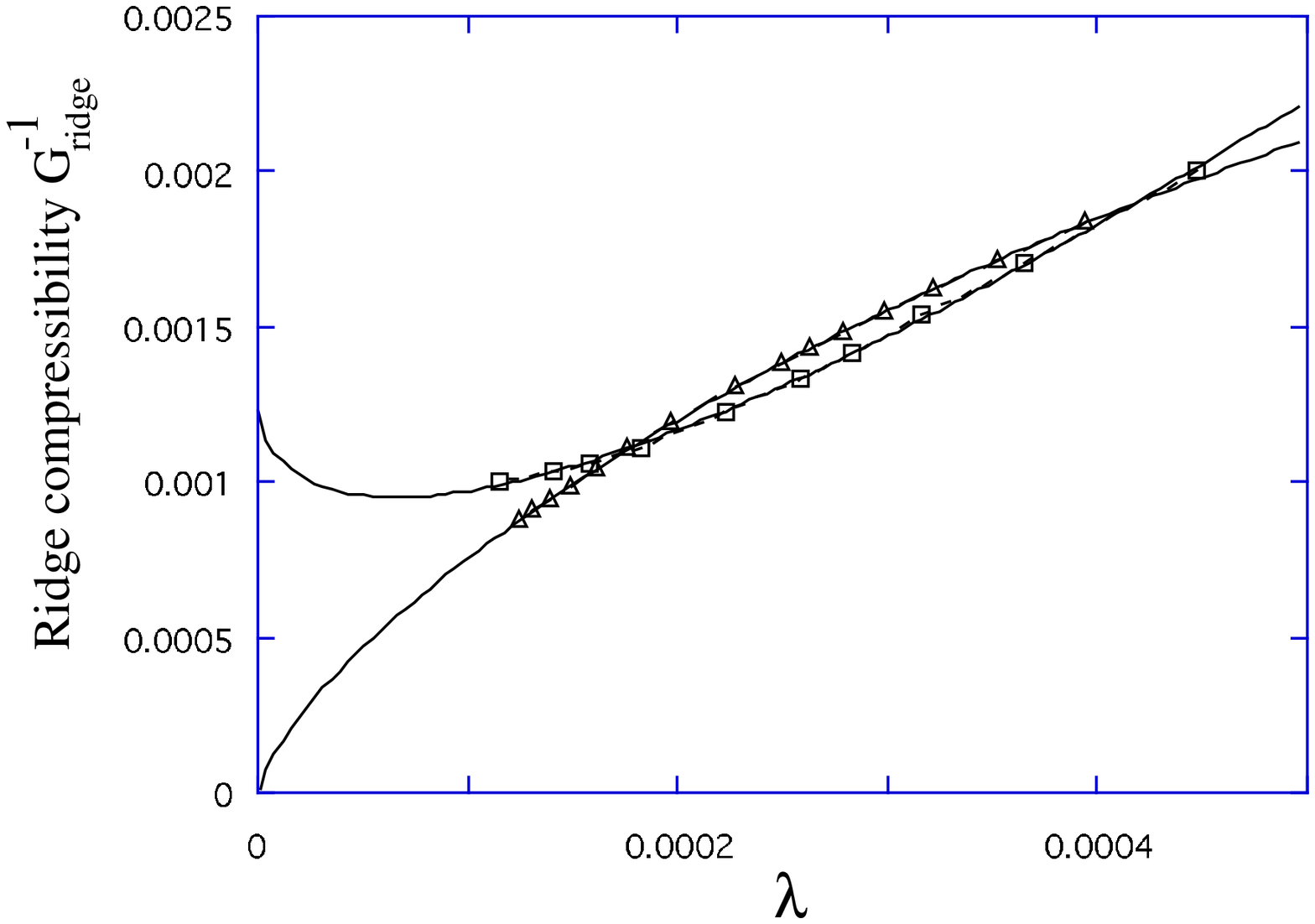}}

FIG.\ 7. The coefficient $G^{-1}_{\rm ridge}$ of the linear $\bar F$
dependence of the ``vertex-to-vertex'' strain created by the
external compressive force $\bar F$ for the strip (squares) and
the tetrahedron (triangles).  Solid lines are fits according
the Eq.\ (\ref{add-strain}).  The tetrahedron
ridge stiffness $G_{\rm ridge}$ is qualitatively greater in the
limit of the vanishing thickness.
\end{figure}

An important question that needs to be settled numerically is
the buckling threshold of the ridge.  
The vertex forces were increased until the shape underwent a radical
change.  This seems to
imply that buckling is a first order phenomenon.
In other words, there are several stable shapes other than
the ridge for a range of compressive forces
below the buckling threshold.  Fig.\ 8 shows
two such buckled bent strip shapes.  A buckled tetrahedron shape shown
in Fig.\ 9 exhibits a more complex buckling pattern.
Shading is proportional to the stretching energy density.
\begin{figure}
\centerline{\epsfxsize=8cm \epsfbox{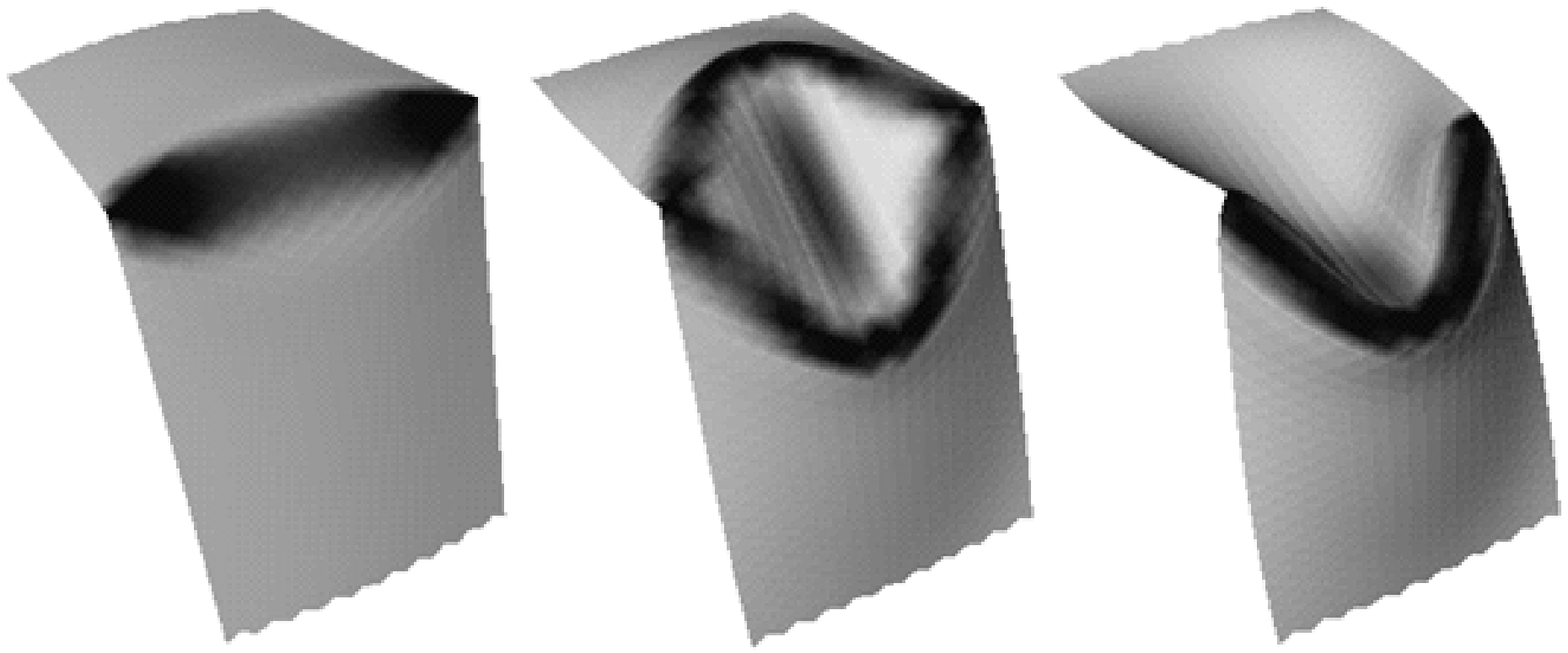}}

FIG.\ 8. A roof-shaped strip (left) is collapsed by corner forces to
obtain two stable buckled shapes as viewed obliquely from above.
The value of the vertex forces must be reduced below the buckling
threshold in order to achieve stability.  Shading is proportional
to the stretching energy density.
\end{figure}
\begin{figure}
\centerline{\epsfxsize=8cm \epsfbox{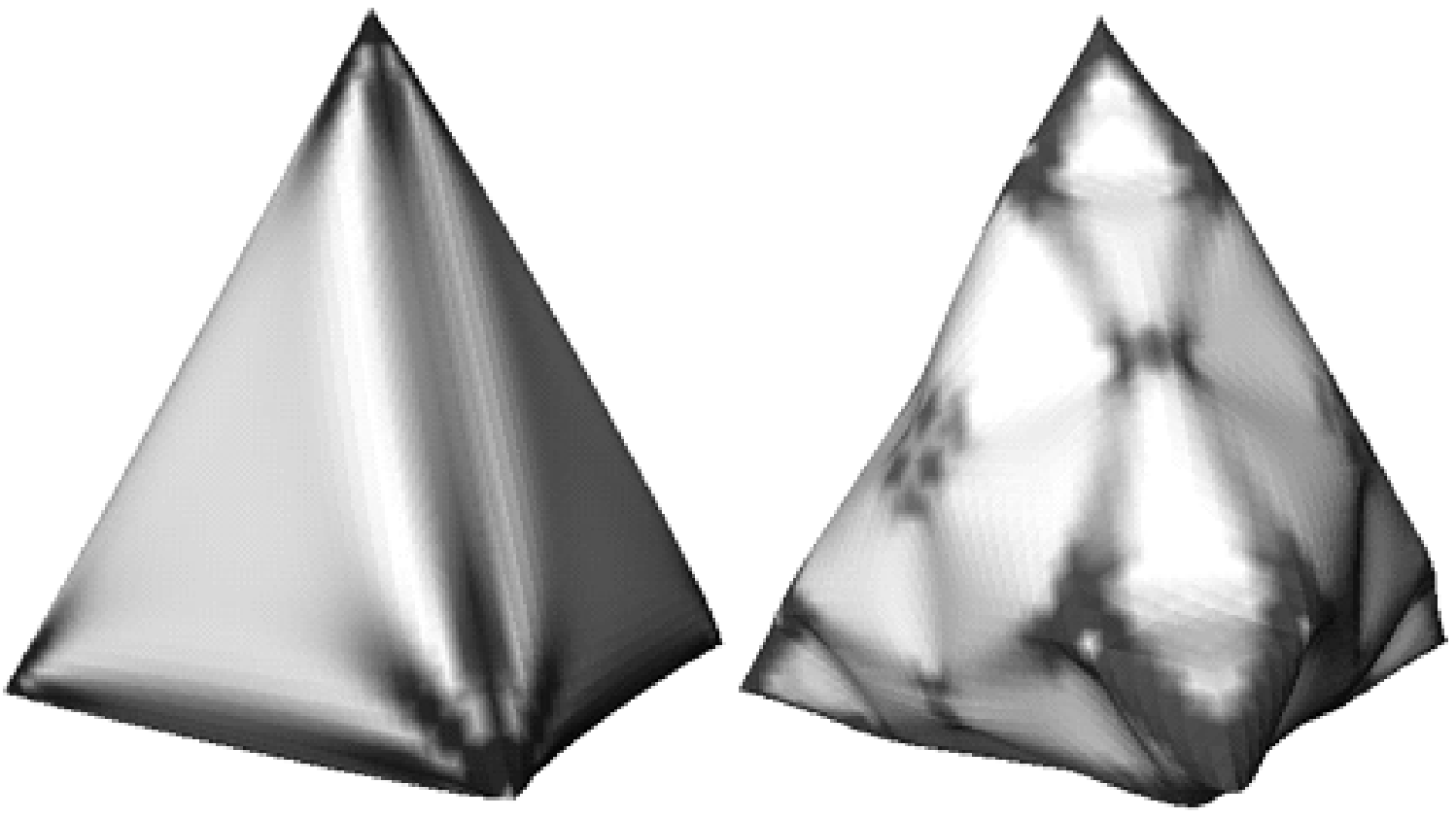}}

FIG.\ 9. A tetrahedron lattice (left) is 
collapsed by application of vertex forces (right).  A complicated 
buckling pattern results.
\end{figure}

In Fig.\ 10 we plot
the ratio of the energy correction to the total elastic
energy of the undisturbed ridge at the bifurcation point
of the loaded tetrahedron.  This ratio seems to be approaching a constant
in the $\lambda \rightarrow 0$ limit which
would agree with the prediction that the energy of a ridge
can only change by a finite fraction before it buckles.
More numerical tests are needed however to establish this
assertion firmly.  We did not attempt to characterize the buckling
of the tetrahedron in further detail.

\begin{figure}
\centerline{\epsfxsize=8cm \epsfbox{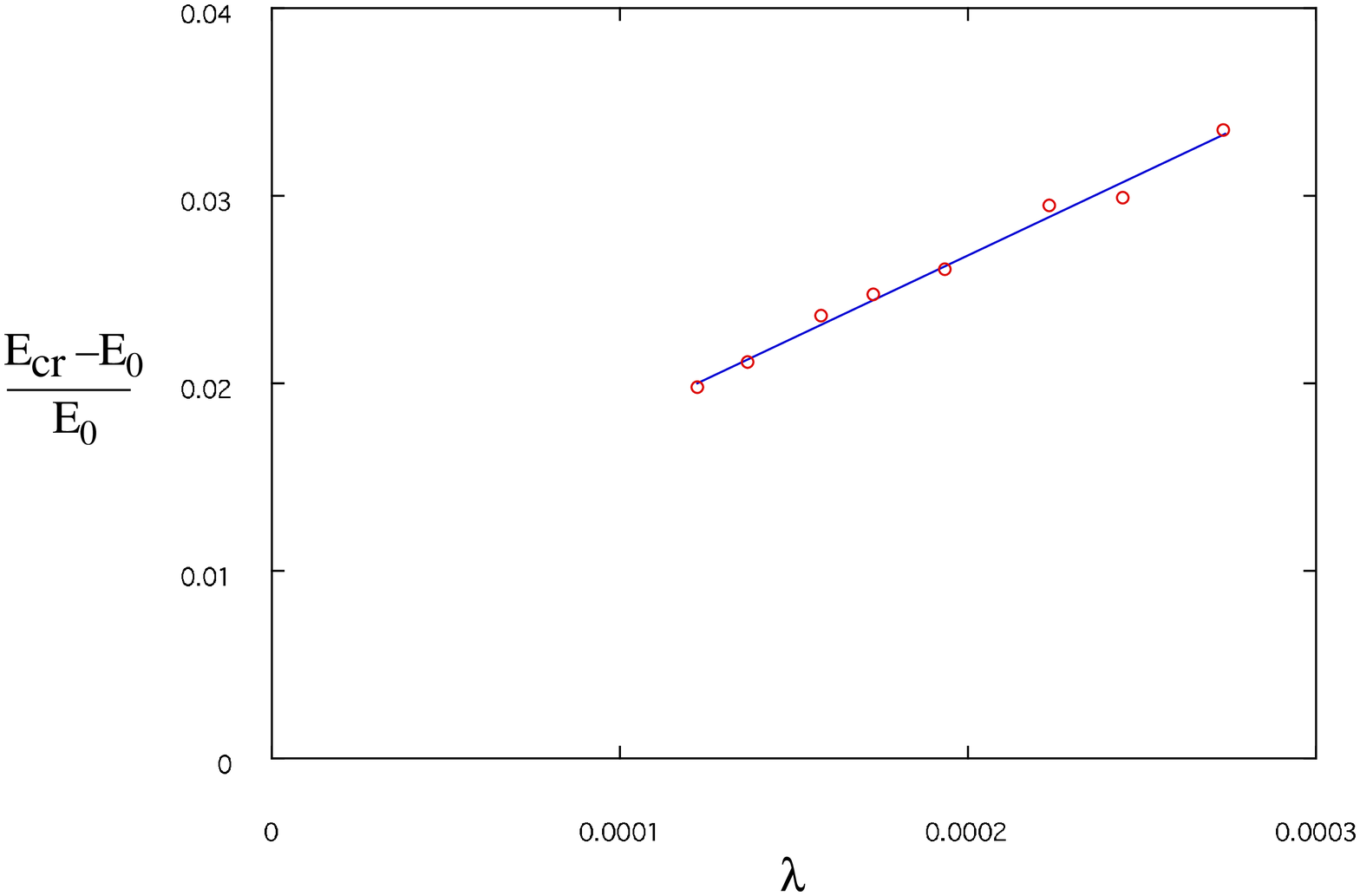}}

FIG.\ 10. Ratio of the energy correction at the buckling threshold
of the tetrahedron of edge length $50b$ to the energy of the
undisturbed tetrahedron as a function of the
dimensionless thickness $\lambda$.  The empirically drawn solid line
suggests that the energy correction varies roughly linearly with $\lambda$.
\end{figure}

The large distance behavior of the ridge ``wings'' was
tested by using a long $52b$ by $500b$ strip.  Fig.\ 11
displays the transverse curvature $C_{yy}$ and the
longitudinal curvature $C_{xx}$
in the units of $X^{-1}$
along a perpendicular bisector of the ridge.
A long almost linear decay of the $C_{xx}$ is evident
whereas $C_{yy}$ decays rapidly to zero.
\begin{figure}
\centerline{\epsfxsize=8cm \epsfbox{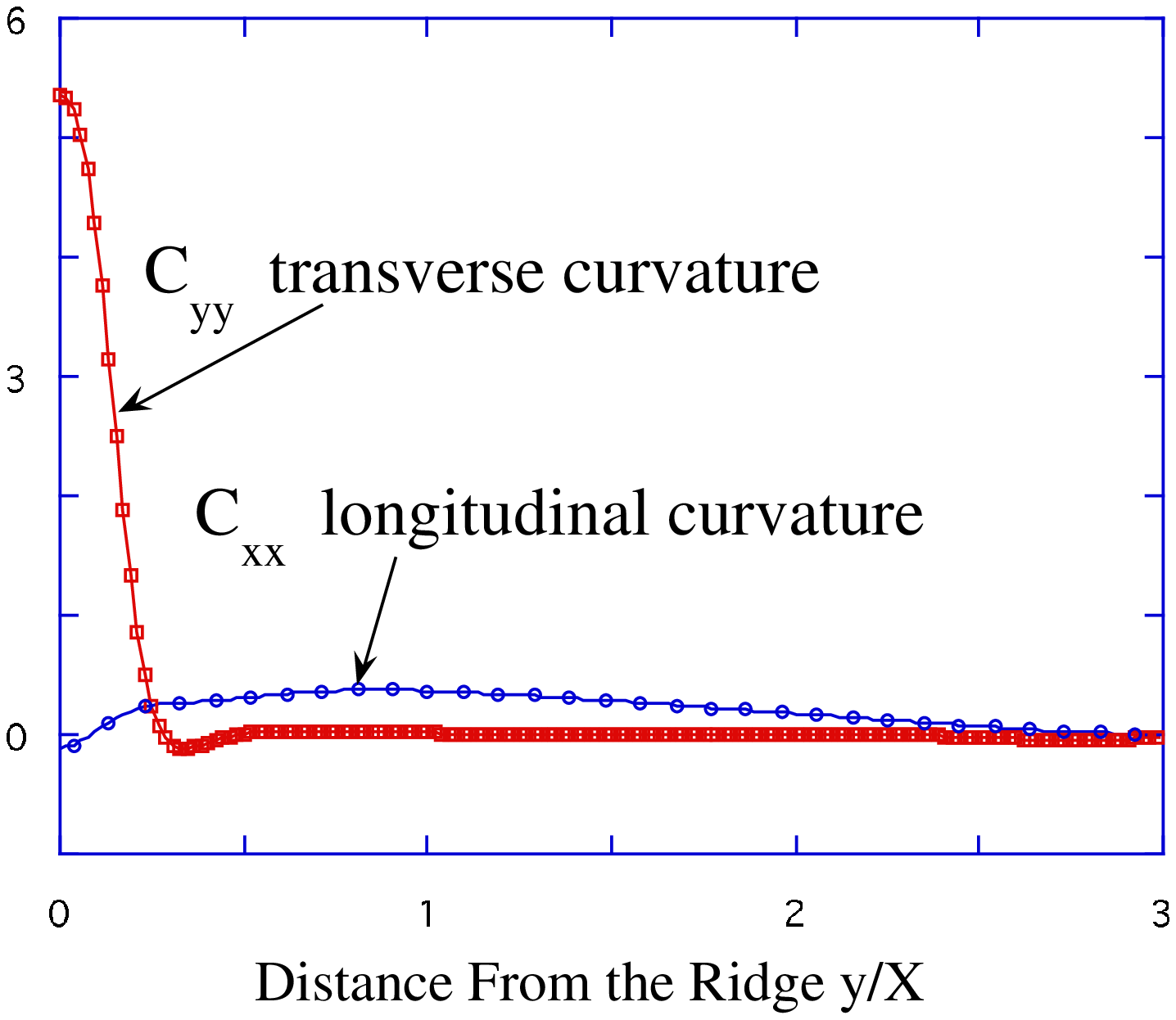}}

FIG.\ 11. Transverse ridge curvature $C_{yy}$ (squares) and
longitudinal curvature $C_{xx}$ (circles) in the units of
$X^{-1}$ along the perpendicular bisector of the ridge vs.
the distance from the ridge for a $52b$ by $500b$ strip.
\end{figure}
We extracted the decay length $L$ for a sequence of ridges
of varying $\lambda$ and plotted the results in Fig.\ 12
versus the predicted scaling variable $\lambda^{-1/3}$.
The linear fit shows that the data agree well with the prediction.

\begin{figure}
\centerline{\epsfxsize=8cm \epsfbox{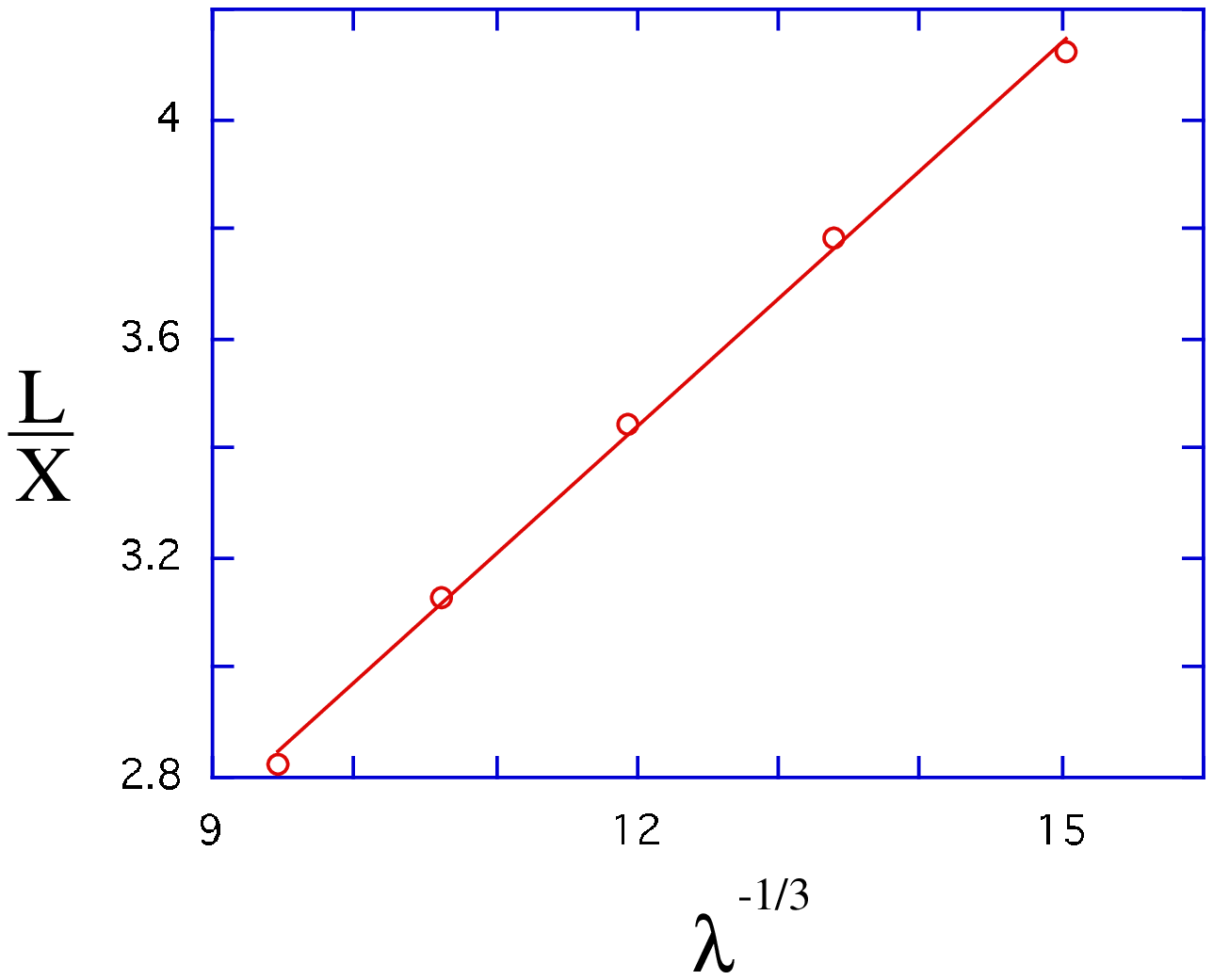}}

FIG.\ 12. Decay length of the longitudinal curvature
$L$ in the units of $X$
vs. the predicted scaling variable $\lambda^{-1/3}$
for a $52b$ by $500b$ strip.
\end{figure}

We next investigated the question of ridge interaction.
Two parallel ridges were created in the strip geometry.
We varied the distance between the ridges $D$ as well
as the thickness $\lambda$.  Detailed interaction features that are
shown in Fig.\ 13 for the $100b$ by $130b$ 
strips for a fixed $\lambda = 0.0002$
depend on the relative orientation of the ridges.  If they are both
concave (or both convex) as seen by looking down on the sheet from above
(configuration U),
they repel at short distances and attract at long distances.
\begin{figure}
\centerline{\epsfxsize=8cm \epsfbox{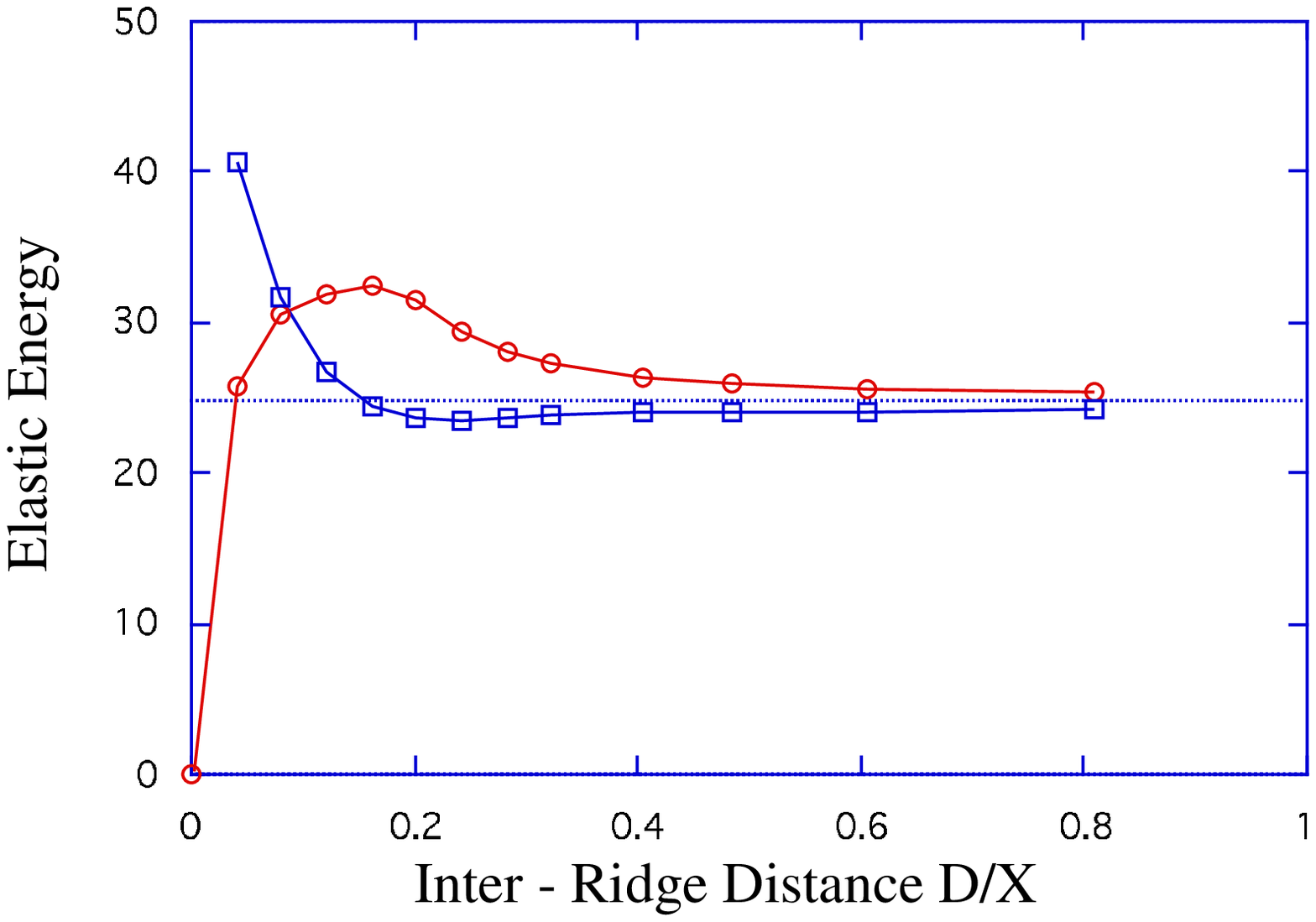}}

FIG.\ 13. Total elastic energy of a two ridge configuration
in the units of $\kappa$ found from a $100b$ by $130b$ strip
bent in two places by $90^\circ$ vs. the distance between the
ridges $D$.  The thickness is $\lambda = 0.0002$.
Square symbols correspond to the ridges
that have the same orientation (configuration 
U), whereas the triangles correspond
to two ridges of the opposite orientation (configuration Z).
\end{figure}
The situation is reversed if the ridges have different orientation
(one is folded up and the other one down, or vice versa: configuration Z).

The signs of the interactions can be readily understood.
At short distances, the two interacting ridges become a
single ridge.  If the two ridges have opposite signs, there is no
ridge when they are brought together and thus the total energy is zero.
If they have the same sign, the combined ridge has twice the dihedral
angle $\alpha$ as the constituent ridge.  Since ridge energy 
scales as $\alpha^{7/3}$ \cite{vonK}, the energy at $D=0$ should be
$2^{7/3}/2$ times that at $D=\infty.$
The behavior of the interaction energy at large $D$
is also understandable.
Here the deformation between the ridges is minimal.
For opposite sign ridges (configuration Z),
the curvature fields created in the region between the ridges have
opposite signs.
For same-sign ridges shaped like a squared letter U, deformations
caused by the two ridges at the midpoint reinforce each other.
In a medium with a quadratic energy functional, such reinforcing
deformations lead to a reduction in energy.
Conversely, two opposite-sign ridges in a Z configuration
have opposing deformations at the midpoint.  This yields a repulsion.

\begin{figure}
\centerline{\epsfxsize=8cm \epsfbox{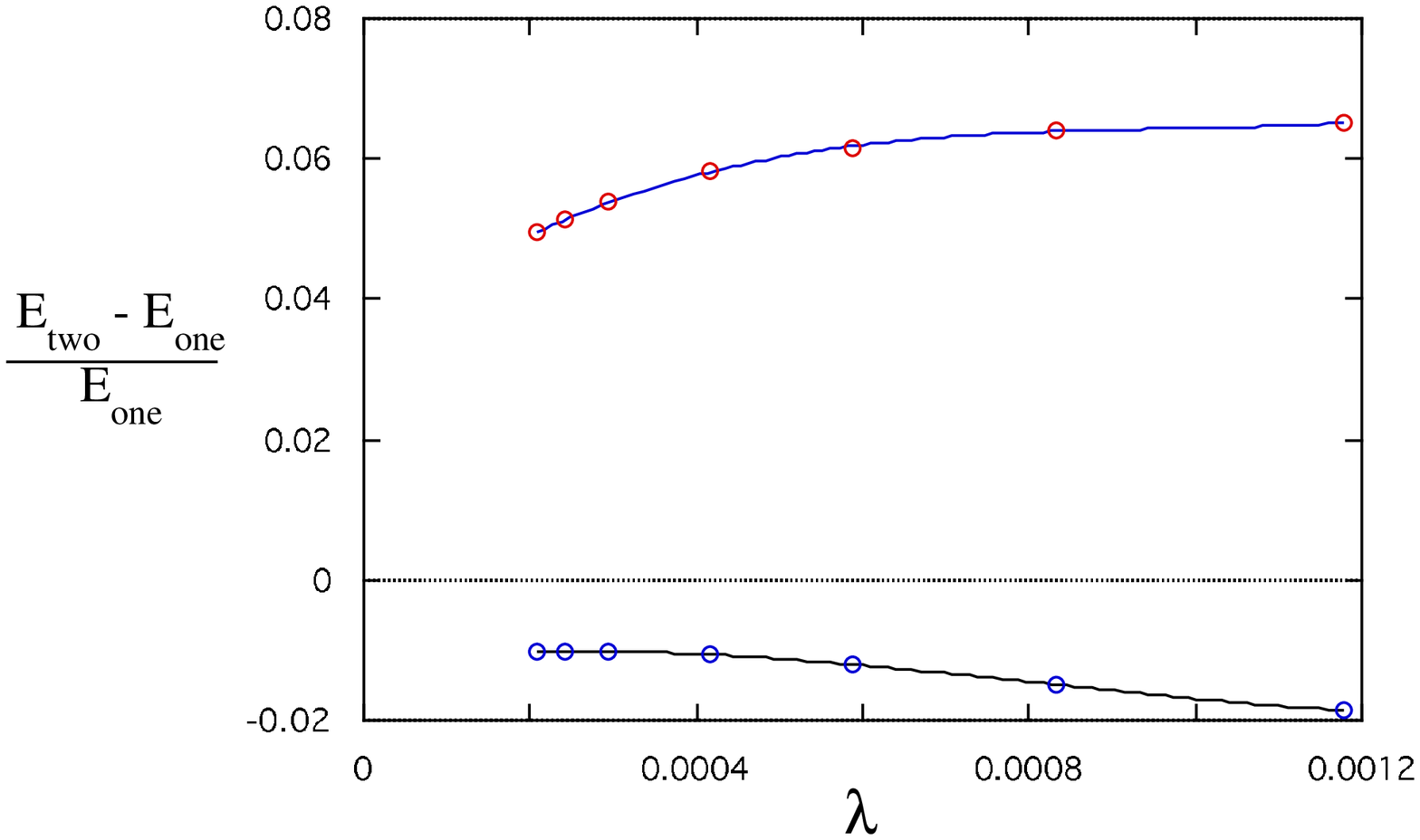}}

FIG.\ 14. Ratio of the difference of the total energy of two
ridges and twice the energy of once ridge $\Delta E$ 
divided by the energy
of one ridge $E$ as a function of $\lambda$ for a fixed inter ridge
separation $D=1.15X$.  The data are obtained from simulation of
a $52b$ by $500b$ strip bent in two places.  Squares correspond to
the configuration U in which the two ridges are have the same
orientation, and triangles correspond to the ridges of
the opposite orientation Z.
\end{figure}

For a complete account of the interaction energy, we must consider
energy stored not only between the ridges, but within each ridge due to
the presence of the other.  To investigate such effects we made a
long $52b$ by $500b$ strip bent by $90^\circ$ in two places. 
This creates two parallel ridges in the
same strip whose relative orientation can be changed.
In Fig.\ 14 we plot the difference of the
elastic energy of the two ridges and twice the energy of one ridge $\Delta E$
divided by the energy of one ridge $E$
as a function $\lambda$ for a fixed
inter-ridge separation of $D=1.15X$.  Squares correspond to
the configuration in which the two ridges have the same
orientation (U), and triangles correspond to the ridges of
the opposite orientation (Z).
The results are consistent with the prediction that this
ratio must approach a finite constant in the $\lambda \rightarrow 0$ limit.
Therefore the prediction that ridge interaction changes the system
by a finite fraction is convincingly confirmed.

\section{Discussion}

In this article we explored properties of the ridge singularity
in thin elastic plates that may be relevant to a quantitative
analysis of crumpled elastic membranes.
A virial theorem that relates the bending and stretching
contributions to the total elastic energy has been derived
from an energy scaling argument and verified numerically.
The virial theorem affords a useful test of elastic energy confinement.
When most of the elastic energy is confined to the ridges
the virial theorem predicts the ratio of the bending 
and stretching energies in the small thickness limit.

We have developed a perturbation expansion scheme that
allows one to calculate the effects of external forces
on the scaling properties of the ridge singularity.
We found that the problem can be decomposed
into first solving for the effect of the
external forces on the flat sheet $\chi_e$ and $f_e$ 
and incorporating this solution into the equations that describe the
ridge singularity in a way that is particularly convenient
for a perturbation expansion.
This method allows one to determine the scaling of the energy 
correction.  This correction scales with the applied
force squared and also a power of thickness that depends
on the details of $\chi_e$ and $f_e$ in the ridge region.
Two different types of scaling were identified.  Generally,
an imposed strain comparable to the strain in a ridge
stores an energy comparable to the ridge energy.
This leads to an effective modulus of order $E_{\rm ridge}/X^{3}
\sim Y \lambda^{8/3}$.  A weaker modulus is possible for
isolated ridges with stress-free boundaries since the imposed
strain can be relaxed in ways not accessible to ridges in a crumpled
sheet.

A feature common to all types of ridge loading is that
the ridge solution becomes unstable when the external
forces change the energy of the ridge by an amount that
is comparable to the original undisturbed ridge energy.
This discovery gives justification to a claim that the energy
of a crumpled elastic sheet can be found 
once the ridge network is characterized
in terms of the ridge sizes $X_i$ and their dihedral
angles $\theta_i$ \cite{vonK}.

We have found that the large distance part of the ridge
solution that matches onto the boundary layer solution
exhibits scaling with the thickness
$\lambda$.  We have established by an energy scaling argument as well
as an extension of the scaling analysis of the \vonK 
equations that small stresses and strains persist
up to a distance $L \sim X\lambda^{-1/3}$ away from the ridge.
Using the framework developed in this paper we found that
the main implication of the ridge ``wings'' is that
ridges located at distances $D \ll L$ away from each
other interact in a way that changes their total
energy by a small but finite fraction.

In the future we propose to build a model for
a quantitative characterization of the ridge network in 
crumpled sheets that incorporates the properties of ridges
uncovered in this article and makes a prediction
for the ridge size distribution.  We hope to characterize the
buckling behavior more thoroughly and verify our hypothesis
that the generic scaling of the modulus is that of the tetrahedron
studied here.

\section*{Acknowledgments}

The authors are grateful to E.~M.~Kramer for many productive and stimulating
discussions.  This work was supported in part by the MRSEC Program
of the National Science Foundation under the award number DMR-9400379
and by the National Science Foundation grant number DMR-9528957.

\end{multicols}
\end{document}